\documentclass[12pt]{article}
\usepackage{amssymb, amsmath,mathtools}
\usepackage[dvips]{graphicx}
\usepackage[sort]{natbib}
\usepackage{epsfig,ifthen,
	amsmath,amssymb,ifthen,
	pstricks,pst-node}
\usepackage{booktabs}
\usepackage{epstopdf}
\DeclareGraphicsExtensions{.ps}
\newcounter{isamac} 
\ifthenelse{\value{isamac}=1}{
\input BoxedEPS          %
\SetTexturesEPSFSpecial  
\HideDisplacementBoxes   %
}{}

\newtheorem{theorem}{Theorem}

\newtheorem{remark}{Remark}

\newcommand{\indm}[2]{\ensuremath{{\mathfrak I}_{\kern-1pt\scriptstyle#1}({\mathcal
#2})}} 

\newcommand{\ind}{\mbox{$\perp \kern-5.5pt \perp$}}

\newcommand{\uned}{\hbox{\kern3pt\raise2.5pt\vbox{\hrule
width9pt height 0.3pt}\kern3pt}}

\newcommand{\dashed}{\hbox{\kern3.05pt\raise2.5pt\vbox{\hrule
width1.7pt height 0.3pt}\kern1.8pt\raise2.5pt\vbox{\hrule
width1.7pt height 0.3pt}\kern1.8pt\raise2.5pt\vbox{\hrule
width1.7pt height 0.3pt}\kern1.8pt\raise2.5pt\vbox{\hrule
width1.7pt height 0.3pt}\kern3.05pt}}

\newcommand{\pedg}[2]{\ensuremath{{\kern0.5pt
\scriptstyle{\ifthenelse{\equal{\head}{#1}}{\lhead\kern0.5pt}{#1\kern0.5pt}}\joinrel\relbar
\negthinspace\relbar\joinrel{\kern0.5pt #2}\kern0.5pt}}}

\newcommand{\pdots}{\hbox{\kern2.5pt\raise1.5pt\hbox{\ensuremath{\ldots}}\ke
rn2.5pt}}  

\numberwithin{equation}{section}
\numberwithin{theorem}{section}
\numberwithin{remark}{section}

\definecolor{mygreen}{RGB}{144,241,47}

\newcommand\given[1][]{\:#1\vert\:}
\newcommand{\talpha}{{\dot{\alpha}}}
\newcommand{\tbeta}{{\dot{\beta}}}

\newcommand{\OR}{odds ratio}
\input{tcilatex}

\usepackage{duerer}

\usepackage{url}
\newcommand\numberthis{\addtocounter{equation}{1}\tag{\theequation}}

\usepackage{times}
\usepackage[left=1in,right=1in,top=1in,bottom=1in]{geometry}

\usepackage{authblk}
\author[1]{Thomas S. Richardson}
\author[2]{James M. Robins}
\author[3]{Linbo Wang\thanks{Address for correspondence: Linbo Wang, Department of Biostatistics, Harvard School of Public Health, 677 Huntington Avenue, Boston, Massachusetts 02115 \\
		Email: linbowang@g.harvard.edu}}
\affil[1]{Department of Statistics, University of Washington, Seattle, WA, USA}
\affil[2]{Department of Epidemiology, Harvard School of Public Health, Boston, MA,
	USA}
\affil[3]{Department of Biostatistics, Harvard School of Public Health, Boston, MA, USA}

	{\raise0.5ex\hbox{$#1$}\! \left/ \! \lower0.5ex\hbox{$#2$}\right.}
	
\begin{document}

	\title{On Modeling and Estimation for the Relative Risk and Risk Difference}

	\date{}
	\clearpage \maketitle
	

\begin{abstract}	
{		A common problem in formulating models for the relative risk
	and risk difference is the variation dependence between these parameters and
	the baseline risk, which is a nuisance model. We address this problem by proposing the conditional log odds-product as a preferred nuisance model.     
	This novel nuisance model facilitates
	maximum-likelihood estimation, but also permits doubly-robust estimation
	for the parameters of interest. Our approach is illustrated via simulations and a data analysis.  An R package {\tt brm} implementing the proposed methods is available on CRAN.}
\end{abstract}	

	\section{Introduction}
	\label{sec:intro}
	
	The odds ratio (OR)  is by far the most common way to model the association between binary random variables.
	The popularity of this measure is driven in part by the ease with which logistic regression can be used to describe
	the way that the odds ratio varies with baseline variables. However, there are fundamental difficulties that arise when attempting to interpret odds ratios. 
	For example, even when treatment is randomized and hence independent of other baseline variables it is well known that the odds ratio is not collapsible, meaning that  the marginal odds ratio will not lie in the convex hull of stratum-specific odds ratios \citep{rothman2008modern}. This non-collapsibility makes it hard to interpret ORs and to compare logistic regression coefficients from different studies. 
	The relative risk (RR) and the risk difference (RD)  are alternative measures that are collapsible and are widely regarded as simpler to interpret. These are defined
	as follows:
	\begin{align*}
	{\rm RR}(v)&= \frac{P(Y\!=\!1\,|\, A\!=\!1,V\!=\! v)}{P(Y\!=\!1\,|\, A\!=\!0,V\!=\! v)},\\[6pt]
	{\rm RD}(v)&= P(Y\!=\!1\,|\, A\!=\!1,V\!=\! v) - P(Y\!=\!1\,|\, A\!=\!0,V\!=\! v),
	\end{align*}
	where $A$ is a binary exposure, $Y$ is a binary response and $V$ is a vector of covariates.
	See \cite{lumley2006relative} for an extensive reference arguing for the use of RR/RD (or monotone transformations thereof) in place of ORs.
	
	Any model for the conditional distribution $P(Y \,|\, A,V)$, such as a logistic regression, may be used to obtain
	an estimate for the RR or RD for a given covariate value $v^*$; see \citet[][pp.439--440]{rothman2008modern},
	\citet{mcnutt2003estimating}.
	Although this approach 
	\emph{does} provide estimates for the RR or RD in any given stratum of $V$, nonetheless the parameters of a logistic model do not directly encode
	the dependence of the RR or RD on $V$. As a consequence the logistic model does not allow one to impose or test parsimonious models for
	the functional form of this dependence, nor does it offer direct insight into 
	which baseline variables are important. 
	
	The current standard approach for modeling the dependence of RR or RD on baseline covariates assumes a Generalized Linear Model (GLM), 
	under which there exist vectors
	$\nu$, $\mu$ such that:
	\begin{equation}
	\label{eqn:glm}
	g\{E[Y|A,V]\} = A \nu^T W + \mu^T\!Z,
	\end{equation}
	where $g(\cdot) $ is the link function, and $W=w(V) $,  $Z=z(V) $ are known vector functions of $V$.
	With $Y$ Bernoulli and 
	$g$  the log link $g(u)\!=\!\log(u)$,  equation \eqref{eqn:glm} represents the mean function induced by a log-binomial (or Poisson) regression model;  when
	$g$ is the identity link  $g(u)\!=\!u$, equation \eqref{eqn:glm} represents the mean function induced by a linear probability (or linear regression) model. 
	
	With these choices for $g(\cdot)$, equation \eqref{eqn:glm} implies, respectively,   that 
	$\log\{\hbox{RR}({V})\} =  \nu^T W$ or
	RD$({V}) =  \nu^T W$. To see this, 
	note that if $g=\log$ then \eqref{eqn:glm} can be rewritten in the 
	following \emph{equivalent} form:
	\begin{flalign}
	\log (\hbox{RR}(V)) = \nu^T W,
	\label{eqn:br1} \\
	\log(p_0(V)) = \mu^T\!Z, \label{eqn:br}
	\end{flalign} 
	where $p_a(V) \equiv E[Y\,|\,A\!=\!a,V]$, $a\!=\!0,1$.
	Similarly with the identity link and RD. 
	Note that \eqref{eqn:br1} models the parameter of interest, while \eqref{eqn:br} is a nuisance model used in estimating the parameter of interest.
	
	However, these models present a new problem: the (joint) range of the left side of \eqref{eqn:br1} and \eqref{eqn:br} is a constrained subspace in $\mathbb{R}^{2}$. That is,  {RR}$(v)$ (or RD$(v)$) and $p_0(v)$ are {\em variation dependent} in the sense that the range of $p_0(v)$ depends on the value of RR$(v)$ (or RD$(v)$). For example, if  for some $v^\dag$, {RR}$(v^\dag) =2$  so that $p_1(v^\dag) = 2\times p_0(v^\dag)$, then clearly $p_0(v^\dag) \leq 0.5$!  Consequently, the range  of $(\log\{\hbox{RR}(v)\},\log\{p_0(v)\})$ or similarly, (RD$(v),p_0(v)$) is strictly smaller than the Cartesian product of their marginal ranges. Although one could easily find smooth monotone transformations of $\log\{\hbox{RR}(v)\}$ (or  RD$(v)$) and $\log\{p_0(v)\}$ (or $p_0(v)$) that map each of their domains individually \emph{onto} $\mathbb{R}$, nonetheless due to the variation dependence, 
	the same issue remains: the joint range of the images of these transformations is still strictly smaller than the Cartesian product of the ranges of these images.
	
\begin{figure}
		\begin{center}
			\begin{minipage}{6cm}
				\begin{center}
					\includegraphics[width=6cm]{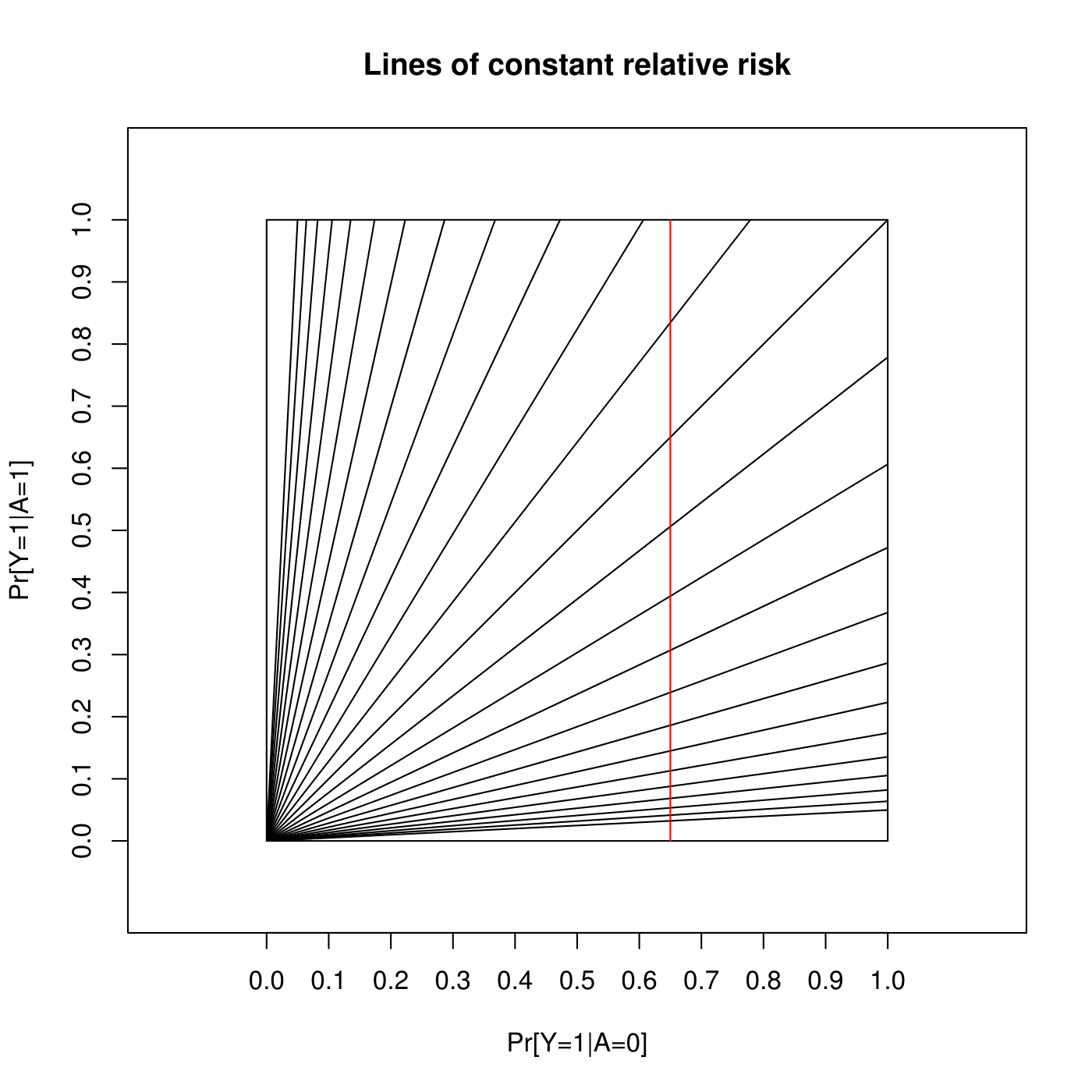}\\
					{\scriptsize(a)}
				\end{center}
			\end{minipage}
			\begin{minipage}{6cm}
				\begin{center}
					\includegraphics[width=6cm]{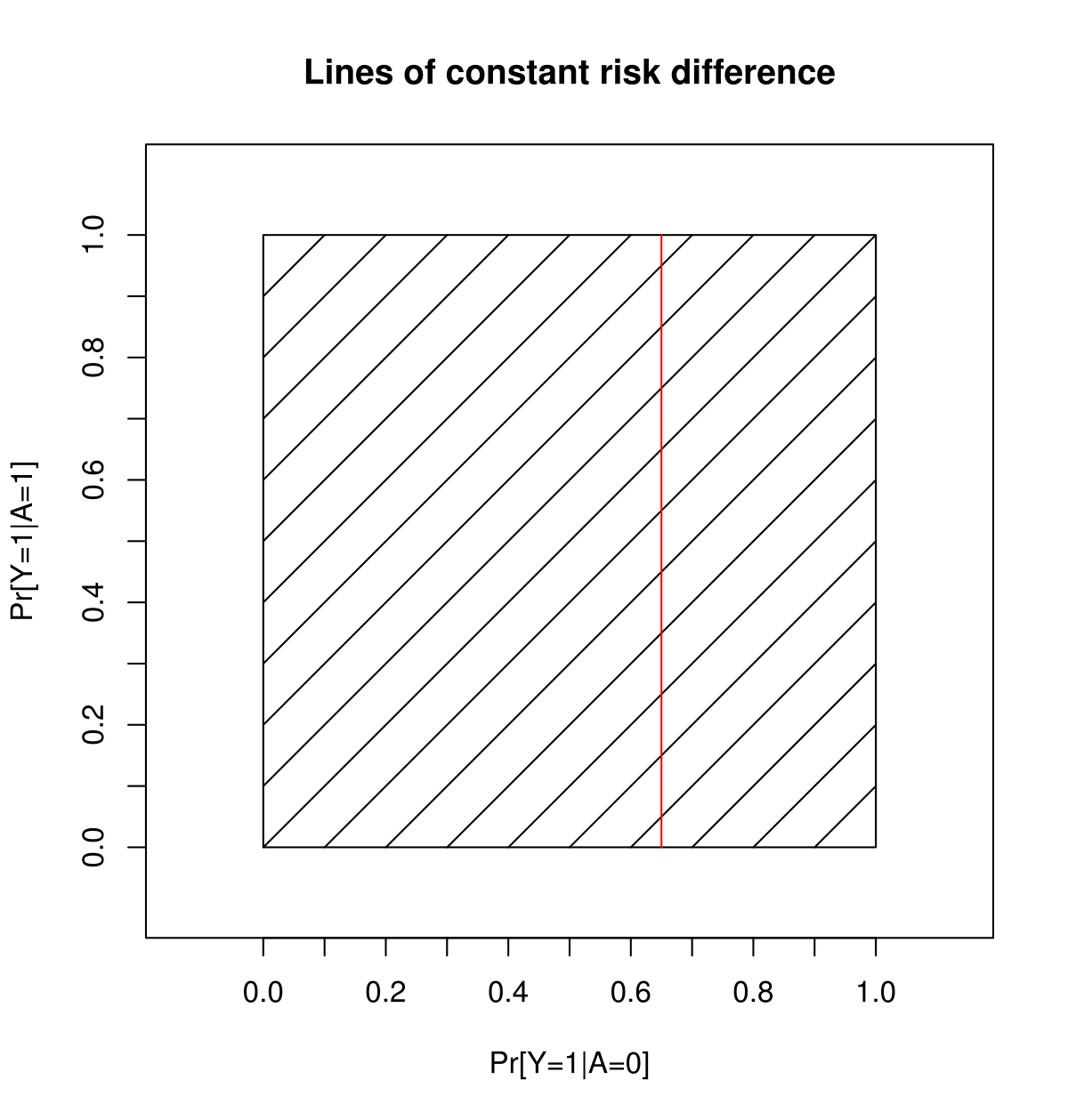}\\
					{\scriptsize(b)}
				\end{center}
			\end{minipage}\\[20pt]
			\begin{minipage}{6cm}
				\begin{center}
					\includegraphics[width=6cm]{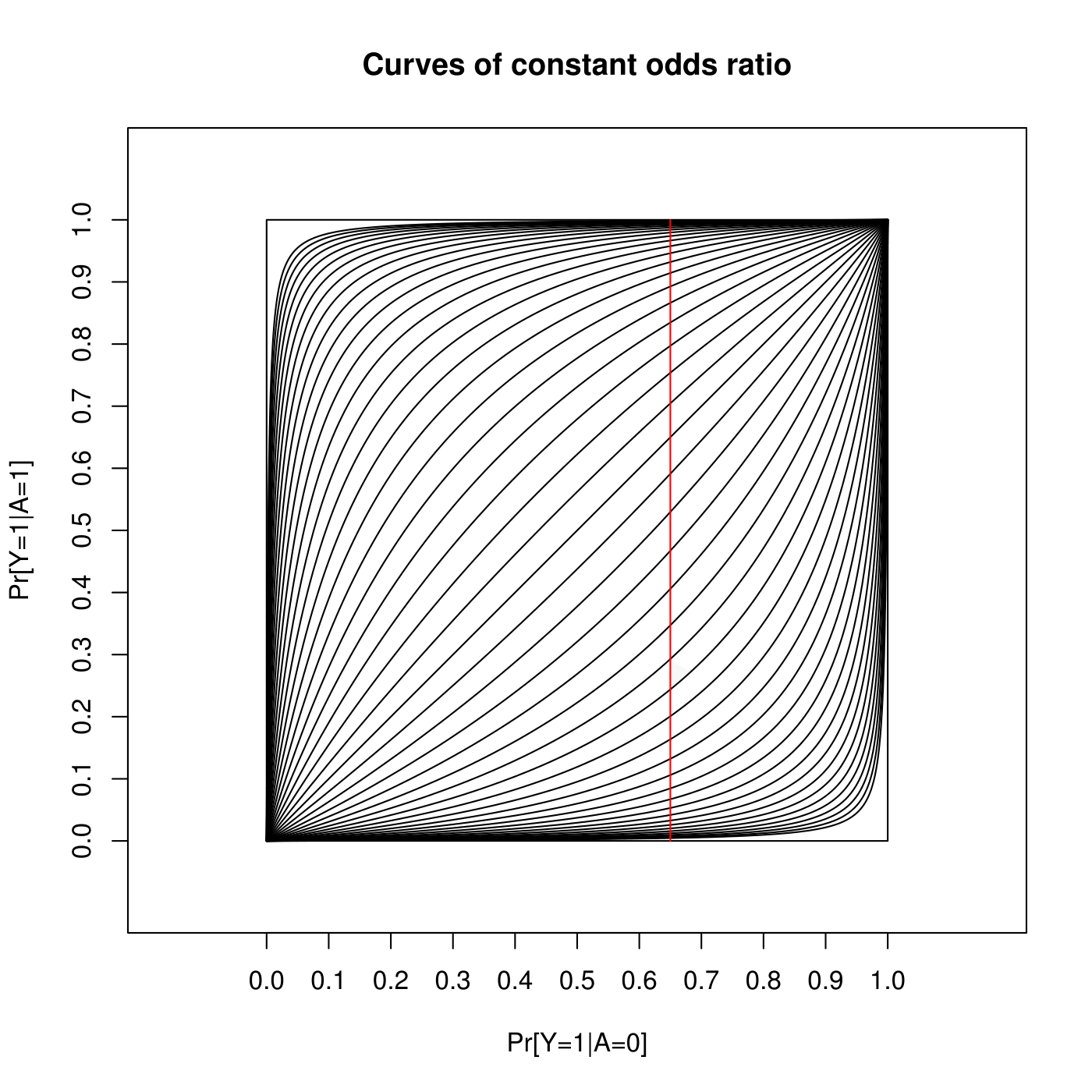}\\
					{\scriptsize(c)}
				\end{center}
			\end{minipage}
			\begin{minipage}{6cm}
				\begin{center}
					\includegraphics[width=6cm]{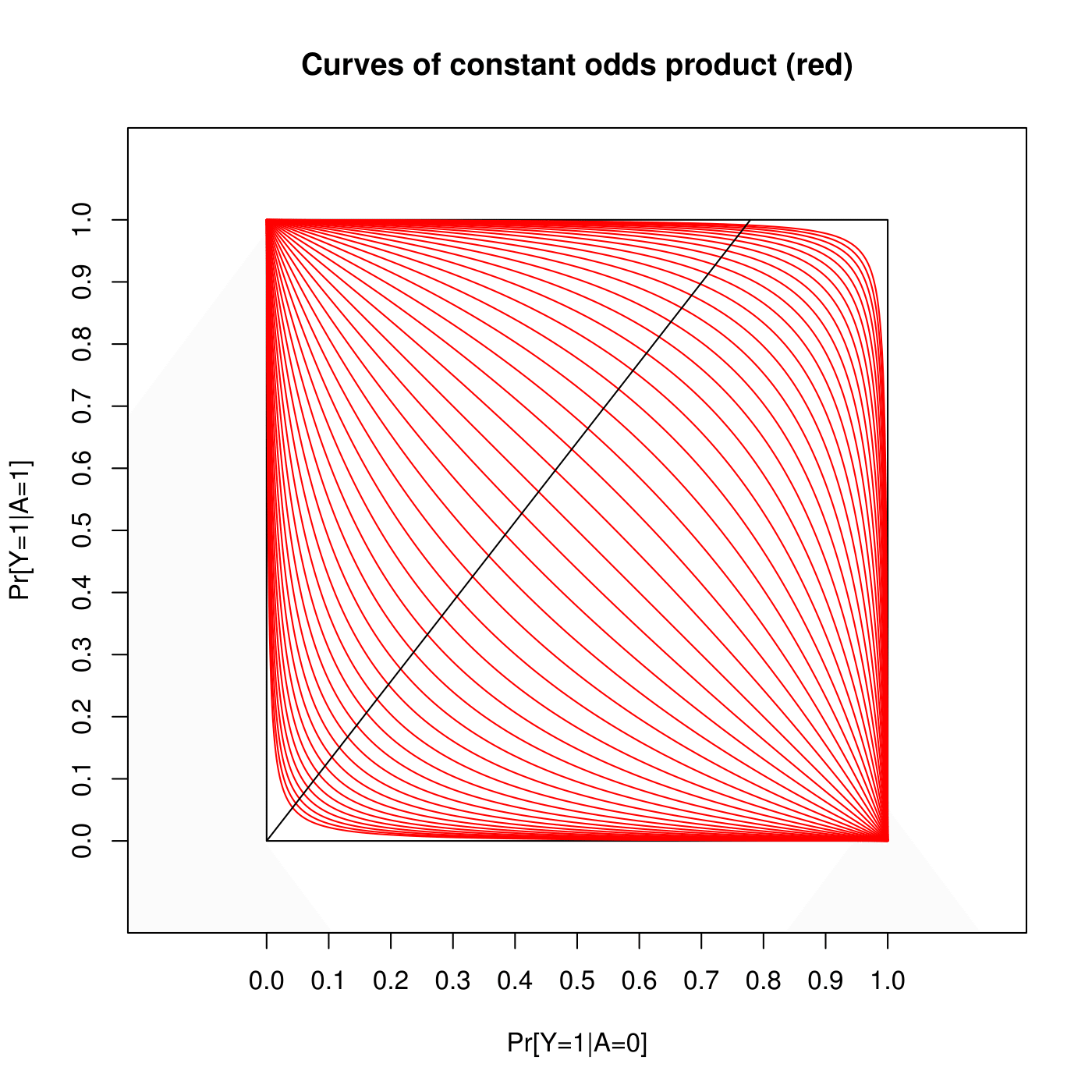}\\
					{\scriptsize(d)}
				\end{center}
			\end{minipage}
		\end{center}
		\caption{L'Abb\'{e} plots: Lines of constant: (a) log RR $\in (-3,-2.75,\ldots,  3)$ (b) RD $\in \{-0.9,-0.8,\ldots,0.9\}$.  (c) Curves of constant log Odds Ratio $\in (-6,-5.75,\ldots,  6)$. The vertical lines in these plots represent a baseline risk of $0.65$ (or a baseline odds of $1.86$).  (d)  Red curves represent specific values of the Odds Product (OP), log OP $\in (-6,-5.75,\ldots,  6)$ and the black line passing through the origin in this plot corresponds to a log RR of $0.25$. 
			\label{fig:one}}
	\end{figure}

	
	{
		\def\OldComma{,}
		\catcode`\,=13
		\def,{%
			\ifmmode%
			\OldComma\discretionary{}{}{}%
			\else%
			\OldComma%
			\fi%
		}%
		In contrast, for logistic regression,  the  OR is variation independent of the nuisance model,  i.e.~log baseline odds: the range of $(\log\{\hbox{\rm OR}(v)\}, \log\{p_0(v)/(1-p_0(v))\})$ is $\mathbb{R}^{2}$. 
		This may be seen directly in Figure \ref{fig:one} (a) - (c). These L'Abb\'{e} plots  show $E[Y\,|\,A=0]$ on the horizontal vs.~$E[Y\,|\,A=1]$ on the vertical \citep{labbe:meta-analysis:1987}.  In plots (a) to (c) the vertical lines represent a specific value for the baseline risk $E[Y\,|\,A\!=\!0]$ (or baseline odds); it may be seen that whereas this vertical line intersects every OR curve in Figure \ref{fig:one}(c), it clearly does not intersect every RR line in (a), nor every RD line in (b).  }
	
	{
		\def\OldComma{,}
		\catcode`\,=13
		\def,{%
			\ifmmode%
			\OldComma\discretionary{}{}{}%
			\else%
			\OldComma%
			\fi%
		}%
		As discussed by many authors, the constraints on the range of $(\log\{\hbox{\rm RR}(v)\}, \log\{p_0(v)\})$ or (RD$(v), p_0(v)$)  have several undesirable consequences:}
	\begin{enumerate}
		\item[{\bf (I)}]  {\bf Models that are always misspecified}: if the support of covariates $(W,Z)$ or parameters ($\nu$,$\mu$) is unbounded, then clearly 
		\eqref{eqn:glm} is misspecified \emph{a priori} owing to the variation dependence. Consequently, any \emph{inference} resulting from \eqref{eqn:glm} is not reliable. 
		\item[{\bf (II)}] {\bf Meaningless predictions}:  even if the supports of both the covariates $(W,Z)$ and the parameters ($\nu$,$\mu$)  are bounded, 
		the estimated $E[Y|A,V]$ applied to a new sample can predict probabilities that lie outside of the range $[0,1]$, and hence be nonsensical.
	\end{enumerate}

	When model \eqref{eqn:glm} is used in combination with maximum likelihood  for parameter estimation (as is often the case in practice), additional critiques apply:
	
	\begin{enumerate}
		\item[{\bf (III)}]  {\bf  Boundary problems}:  the maximum often lies on the boundary of the constrained parameter space. 	This may give rise to computational difficulties in fitting these models. For example,	standard statistical software may report failed convergence
		when attempting to fit log-binomial models in certain settings \citep{lumley2006relative,mcnutt2003estimating,williamson2013log}. 
		\item[{\bf (IV)}]  {\bf Lack of robustness to misspecification of nuisance models}:  even when the model \eqref{eqn:br1}  for  log RR or RD  is correct, misspecification of  \textit{the functional form } ${\mu^{T}\!Z}$ for the log baseline risk or 
		baseline risk generally results in inconsistent estimation of the
		parameter vector $\nu $. This is particularly problematic since, although the log baseline risk or the baseline risk model is required for estimating the RR or RD via maximum likelihood with a GLM \eqref{eqn:glm},  it is not the target of inference.
	\end{enumerate}

	We note that although problems (I) - (III) exist with the log-binomial regression/linear probability models these problems do not arise with the logistic and certain other GLMs. However, problem (IV) is shared by MLEs from all GLMs, including logistic regression.
	
	Owing to problems (I) - (III), applied researchers often choose to  fit logistic regressions and estimate the OR instead of the RR or RD. 
	Arguably this corresponds to choosing the parameter of interest to make it variation independent of parameters of the simplest nuisance model, i.e.~the baseline risk $p_0(v)$.  However, as argued by many authors  \citep[e.g.,][]{hellevik2009linear}, the choice of statistical model should target the parameter of substantive interest
	rather than the other way around. Thus our  approach is to develop an unconstrained nuisance model $\phi(v)$ that is variation independent of RR$(v)$ or RD$(v)$.
	In this way, we solve problems (I) - (III) without changing the parameter of interest.

	
	On the other hand even with the novel nuisance model, maximum likelihood estimation still suffers from problem (IV).  
	To help alleviate this we propose \emph{doubly-robust} estimators of models for (monotone transforms of) the RR and RD that are consistent and asymptotically normal (CAN) even when the nuisance model $\phi({v})$ is misspecified, provided that we have a correctly specified model for the exposure probability $P(A\!=\!1\,|\,{V=v})$, also known as the {\em propensity score}. 
	In contrast, it is not possible to construct such a doubly-robust estimator for the \OR \ \citep[][Theorem 3]{tchetgen2010doubly}. Note that this is of practical importance since in a designed experiment the propensity score will be known by construction.

	The rest of this article is organized as follows. In Section \ref{sec:bivariate_link} we propose our novel nuisance model $\phi(v)$ and briefly outline 
	maximum likelihood estimation. 
	In Section \ref{sec:semi-parametric}, we describe a doubly robust semi-parametric approach to estimation.  Sections \ref{sec:simulations}
	and \ref{sec:data-analysis} contain simulations and an illustrative data analysis.  Section \ref{sec:discuss} contains a summary.

\section{Nuisance model specification}
\label{sec:bivariate_link}

To solve problems (I) -- (III),  the nuisance model $\phi(v)$ needs to be unconstrained and variation independent of (a suitable monotone transformation of) the parameter of interest, $\theta(v)$. Here $\theta(v)$ is a monotone transformation of RR$(v)$ or RD$(v)$ that maps the domain of $V$ onto $\mathbb{R}$. Specifically, our choice for $\theta(v)$ is either $\log\{\hbox{\rm RR}(v)\}$ or arctanh$\{\hbox{\rm RD}(v)\}$ ($=\log\{(1+\hbox{\rm RD}(v))/(1-\hbox{\rm RD}(v))\}$), and our choice for $\phi(v)$ is  the $v$-specific log Odds Product (OP):
\begin{equation*}
\phi \left(v\right) \equiv \log {\dfrac{p_0(v) p_1(v)}{(1-p_0(v))(1-p_1(v))}}. 
\end{equation*}
This approach is illustrated in Figure \ref{fig:one}(d), where each curve corresponds to a specific value of OP.
One can see that  $\log(\hbox{\rm OP})$ ranges from $-\infty$ to $\infty$, and
each contour line of OP intersects every RR line (see Fig.~\ref{fig:one}(a)) and every RD line (see Fig.~\ref{fig:one}(b))
in exactly one point. 
Hence if we specify a parametric form for $(\theta(v),\phi(v))$ such that
\begin{eqnarray}
\theta \left(V\right)  &=&\alpha^{T}W  \label{bisemi}, \\
\phi \left(V\right)  &=&\beta^{T}\!Z \label{bibase},
\end{eqnarray}%
then the parameter of interest, $\alpha$,  and the nuisance parameter, $\beta$, are variation independent.

%


For identification and estimation of the parameters $(\alpha,\beta)$, we note that if we choose the log odds product as $\phi(v) $, then the map given by
\begin{equation}
\left(\theta(v), \phi(v) \right) \rightarrow
\left(P(Y\!=\!1|A\!=\!0,V=v),P(Y\!=\!1|A\!=\!1,V=v)\right),   \label{eqn:bilink}
\end{equation}
is a smooth
bijection from $\mathbb{R}^{2}$ to $(0,1)\times (0,1)$.  In fact, 
the inverse map is given
in closed form by
solving quadratic equations. Specifically, for $\theta(v) = \log \hbox{RR}(v)$, we have
%
\begin{flalign*}
& p_{0}(v) \equiv p(Y=1 \mid A=0,V=v)\\[8pt]
&\quad=    \dfrac{-(e^{\theta (v)}+1)e^{\phi (v)}+ 
	\sqrt{e^{2\phi(v)}(e^{\theta (v)}+1)^2 + 4 e^{\theta(v)+\phi(v)}(1-e^{\phi(v)})}}
{2 e^{\theta(v)}(1-e^{\phi(v)})},  \numberthis \label{eqn:p0V_RR} \\[12pt]
& p_{1}(v) \equiv p(Y=1 \mid A=1,V=v)= p_{0}(v) e^{\theta(v)};
\end{flalign*}
similarly for $\theta(v) =$ arctanh$(\hbox{RD}(v))$ we have
%
\begin{flalign*}
& p_{0}(v) \equiv p(Y=1 \mid A=0,V=v)\\[8pt]
&\quad=  \dfrac{e^{\phi(v)}(2-\rho(v))+\rho(v)-
	\sqrt{\{e^{\phi(v)}(\rho(v)-2)-\rho(v)\}^2+4e^{\phi(v)}(1-\rho(v))(1-e^{\phi(v)})}}{2(e^{\phi(v)}-1)}, \numberthis \label{eqn:p0V_RD}  \\[12pt]
& p_{1}(v) \equiv p(Y=1 \mid A=1,V=v)= p_{0}(v) + \rho(v),
\end{flalign*}
where $\rho(v)\equiv \tanh (\theta(v))$.

\begin{figure}
	\begin{center}
		\begin{minipage}{0.48\textwidth}
			\begin{center}
				\includegraphics[width=\textwidth]{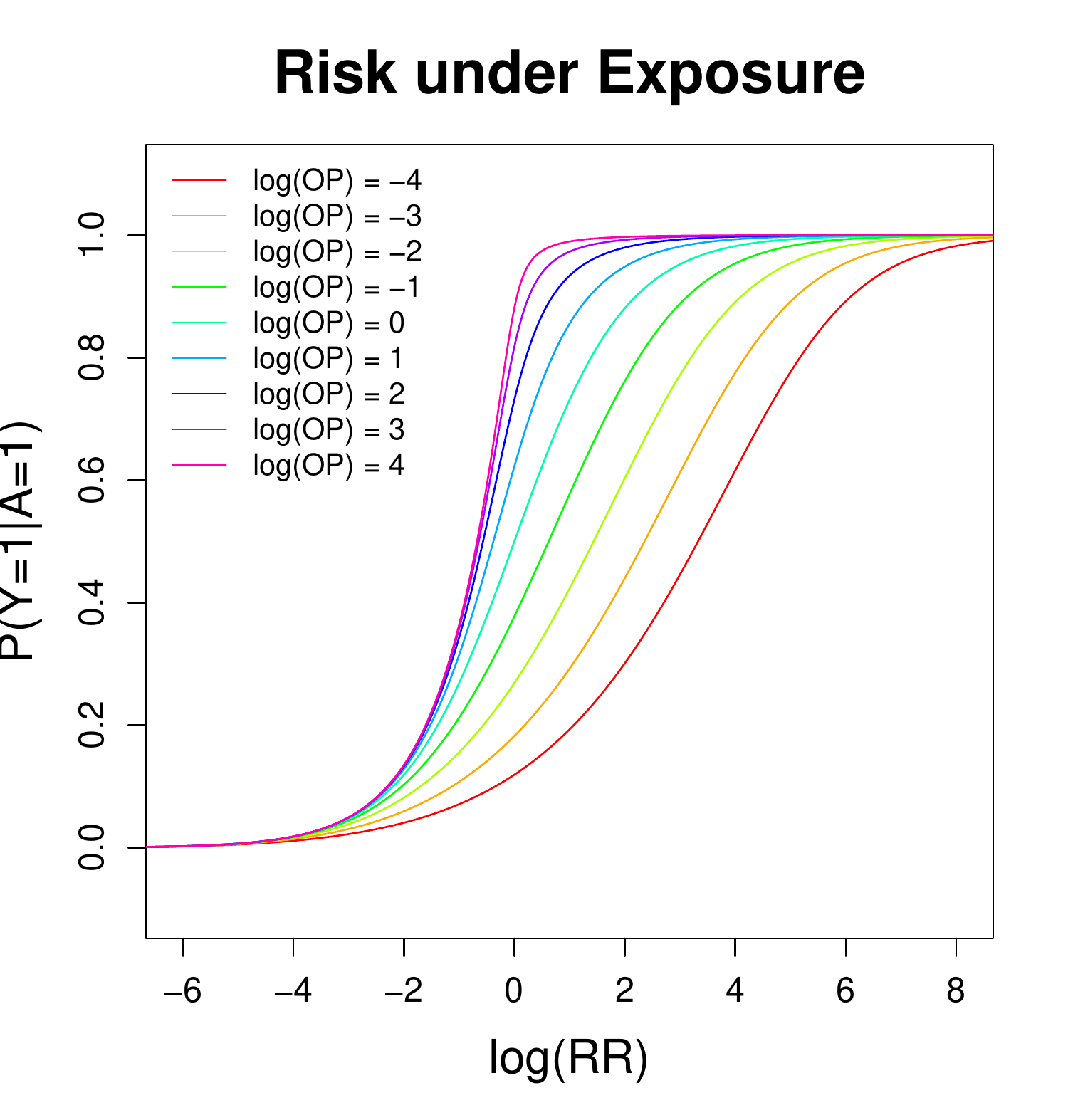}\\
			\end{center}
		\end{minipage}
		\begin{minipage}{0.48\textwidth}
			\begin{center}
				\includegraphics[width=\textwidth]{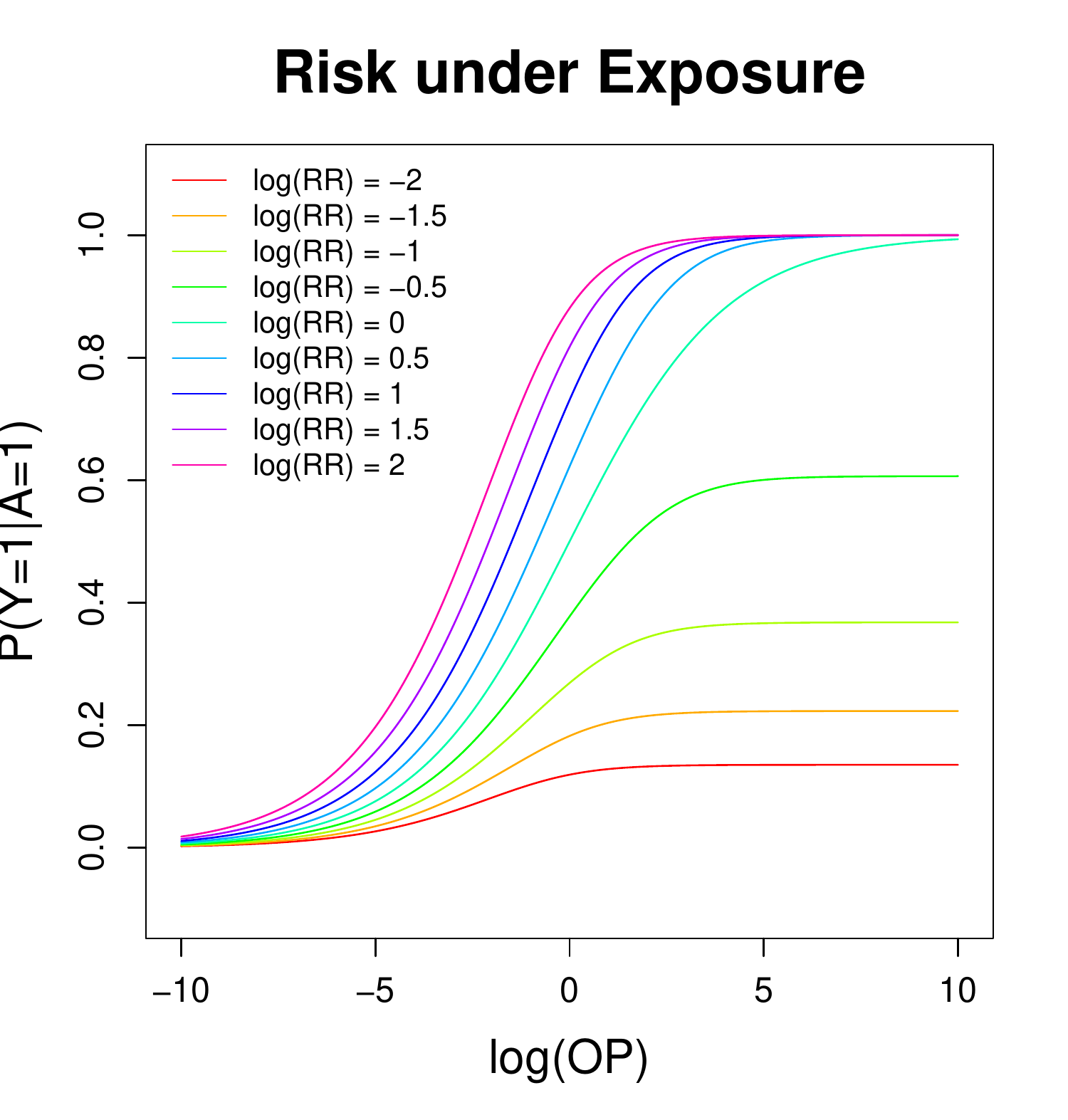}\\
			\end{center}
		\end{minipage}
		\begin{minipage}{0.48\textwidth}
			\begin{center}
				\includegraphics[width=\textwidth]{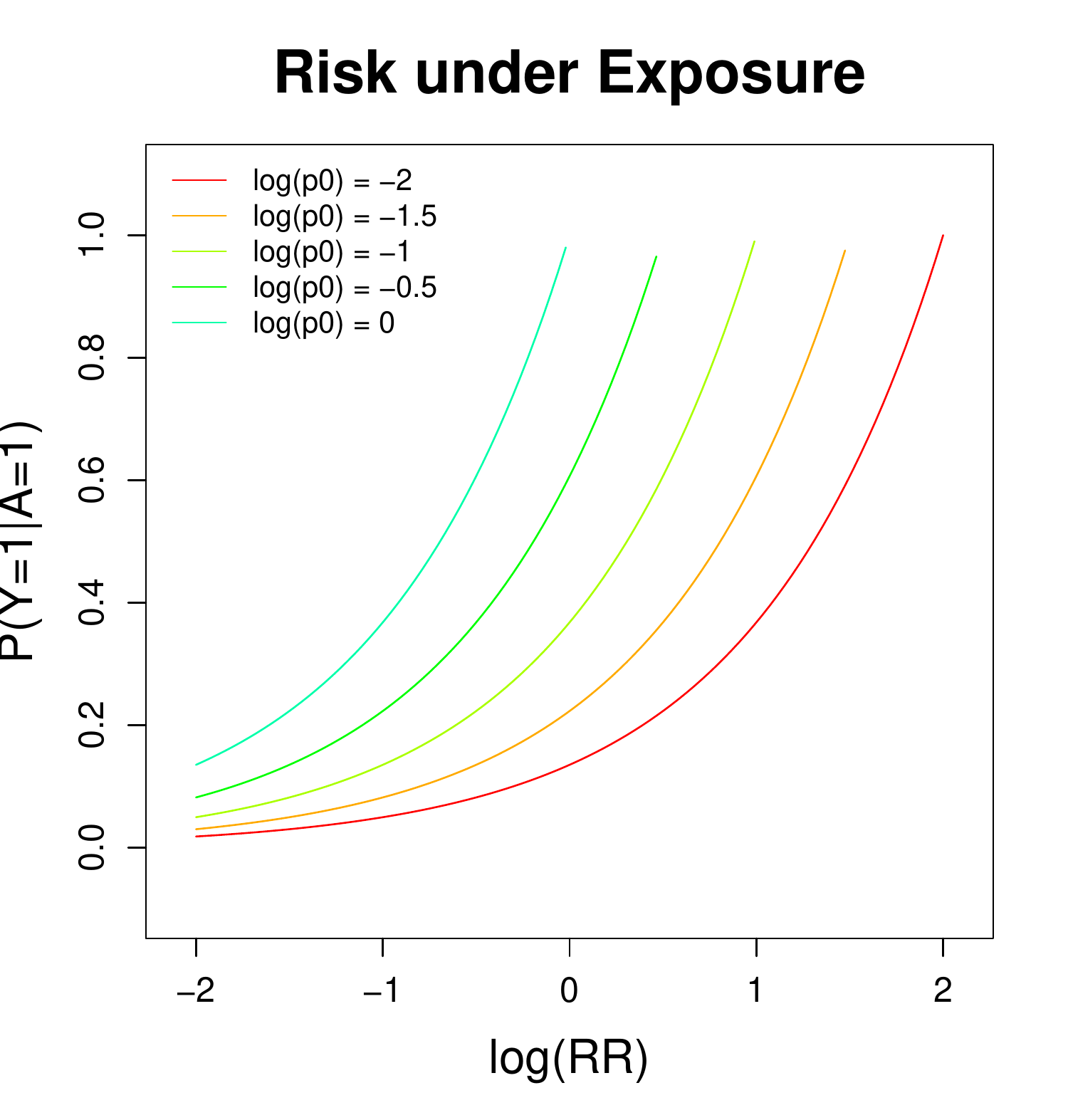}\\
			\end{center}
		\end{minipage}
		\begin{minipage}{0.48\textwidth}
			\begin{center}
				\includegraphics[width=\textwidth]{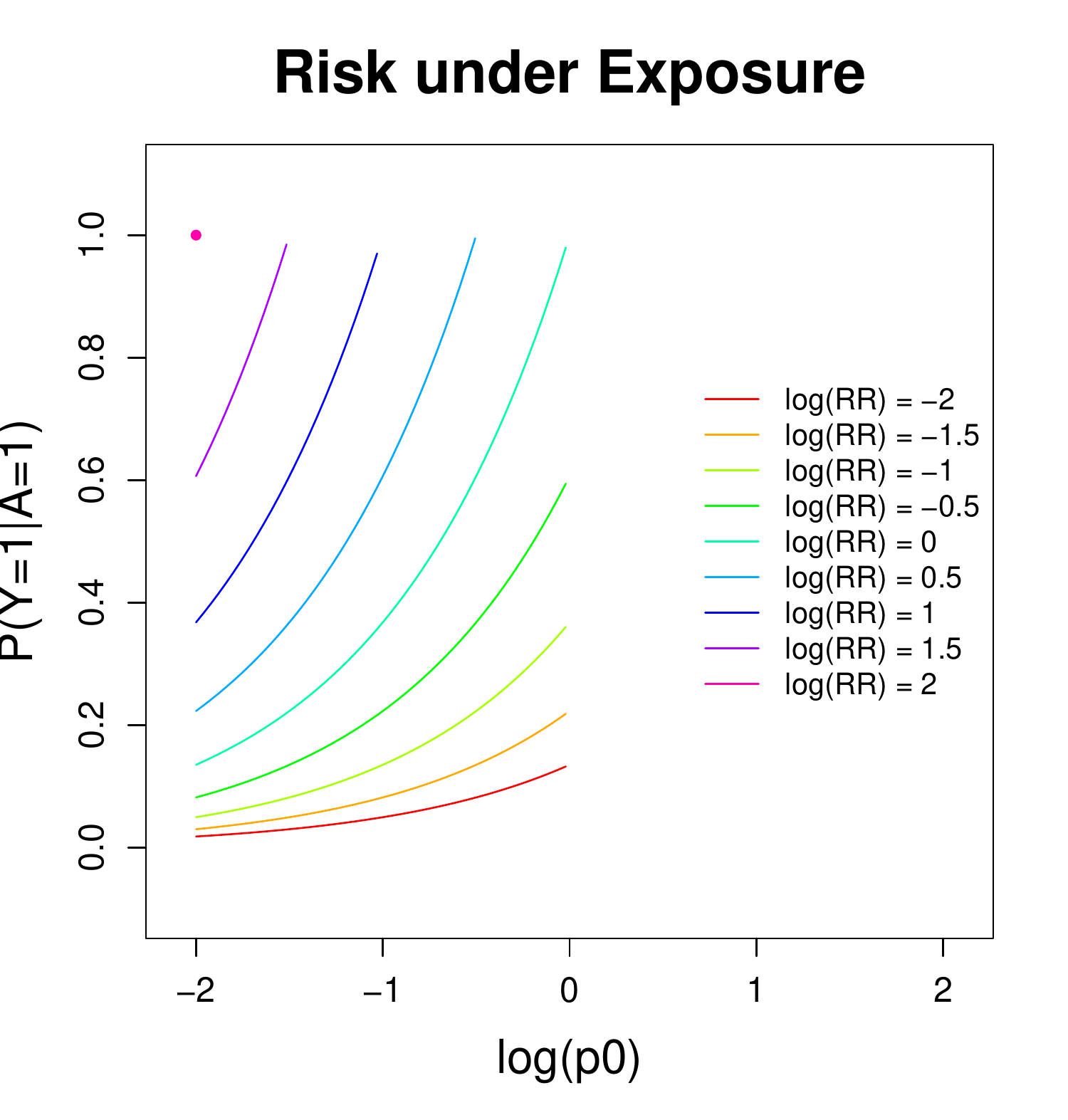}\\
			\end{center}
		\end{minipage}
	\end{center}
	\caption{Implied probabilities from the proposed relative risk model (upper panels) and the Poisson mean model (lower panels). For the proposed relative risk model, curves for implied risks under non-exposure have the same shapes to those for implied risk under exposure.}
	\label{fig:fitted_probabilities}
\end{figure}

As an illustration, Figure \ref{fig:fitted_probabilities} plots $p_1(v)$ as functions of $\theta(v)$ and $\phi(v)$, where $\theta(v) = \log \text{RR}(v)$.  Plots with $p_0(v)$ are similar in shapes and hence omitted; plots with risk difference are given in the on-line supplementary materials. Under the proposed models,  the inverse maps \eqref{eqn:p0V_RR} and \eqref{eqn:p0V_RD} are sigmoid functions of $\theta(v)$ and $\phi(v)$; these are similar to the case with a logistic regression model. In contrast, as one can see from the lower panels of Figure \ref{fig:fitted_probabilities}, the inverse maps from a Poisson mean model do not level off even when the probabilities approach 1.

Under the model given by ({\ref{bisemi}) and (\ref{bibase}), the parameters $\alpha$  and $\beta$ may be estimated directly via maximum likelihood without the need for constrained maximization.
	The log-likelihood for one observation is given by: 
	\begin{eqnarray*}
		\ell(\alpha, \beta)  &=&Y\log \left({P(Y\!=\!1|A,V;\alpha, \beta) }%
		\right) +(1-Y)\log \left({P(Y\!=\!0|A,V;\alpha, \beta) }\right).
	\end{eqnarray*}%
	%
	%
	Likelihood-based confidence intervals for $\alpha$ and $\beta%
	$ can then be obtained in standard fashion.


	We close this section with several remarks. First, there are other nuisance parameters that are also unconstrained and variation independent of the (log) relative risk. For example \cite{tchetgen2013estimation} considers 
	an alternative nuisance parameter, $\tilde{\phi}(V) = \log(P(Y=1|A=0,V) / P(Y=1|A=0))$, for estimating the relative risk. 
	Similar to the log odds product, $\tilde{\phi}(V)$ is also unconstrained and variation independent of the (log) relative
	risk. However, $\tilde{\phi}(V)$ does not give rise to the full likelihood.
	In other words, with $\tilde{\phi}(V)$, equation \eqref{eqn:bilink}  is no longer a bijection from $\mathbb{R}^{2}$ to $(0,1)\times (0,1)$. Consequently, one cannot make
	predictions or use maximum likelihood for parameter estimation.
	%
	
	There are clearly many choices for $\phi(v) $ that would make the bivariate
	map \eqref{eqn:bilink} a smooth bijection. However, the log odds product has the advantage that it is simple and the inverse map can be obtained in a closed form.  It is worth noting that  
	the odds product is used solely as the nuisance model to form a parameterization in conjunction with (a suitable monotone transformation of) the RR or RD, which is the target of inference. Thus the fact that the odds product is nonlinear and hence not collapsible  does not lead to problems of interpretation for the RR$(v)$ or RD$(v)$. 
	Furthermore, formulating a model for the log odds product is similar in spirit to the specification of a logistic regression model. In fact, suppose one specifies a logistic model for the variable $\tilde{Y} \triangleq AY + (1-A) (1-Y)$, such as $\text{logit}(P(\tilde{Y}=1|A,V)) = \tilde{\alpha}^T W + \beta^T AZ,$ then this implies a model for the log odds product (of outcome $Y$): 
	$\log \text{OP}(V) = \beta^T Z$. Note here $\tilde{Y}$ is an indicator function that the observed outcome $Y$ ``agrees'' with the treatment indicator $A$.
	Consideration of $\tilde{Y}$ here is somewhat similar in spirit to the `switch relative risk' \citep{van2007estimation}. 
	
Finally, in the next section we address problem (IV) by introducing a doubly robust estimating procedure that provides an extra layer of protection against misspecification of the log odds product model.

	\section{Semiparametric estimation}\label{sec:semi-parametric}
	
	In the remainder of the paper we will always take model \eqref{bisemi} to be true, so that $\alpha$ is well-defined.
	If the nuisance model (\ref{bibase}) is misspecified, 
	the maximum likelihood estimate for $\alpha$ may not be consistent.
	To address this we propose an estimator that is consistent for $\alpha$, 
	even if (\ref{bibase}) is incorrect, provided that we have correctly specified a model for the 
	\textit{propensity score}: 
	\begin{equation}\label{eq:propensity-score}
	e(V) \equiv P(A=1\,|\,V).
	\end{equation}
	In more detail, we construct a locally semiparametric efficient estimator and show that it is  CAN if
	either (but not necessarily both) model \eqref{bibase} \textit{or} \eqref{eq:propensity-score} is correct. Such estimators are commonly referred to as \emph{doubly robust} \citep{robins2001comments,van2003unified}. 
	
	The following theorem provides the basis for our estimator.  Throughout this section, we make the positivity assumption that with probability $1$, 
	$e(V), p_0(V), p_1(V) \in (0,1)$.
	
	\begin{theorem}
		\label{thm:efficient_score}
		Let $\talpha$ be the unknown true value of $\alpha$, and $\dot{\eta}$ be the unknown true value of an infinite dimensional nuisance parameter consisting of $
		\phi(\cdot), e(\cdot)$ and the marginal distribution of $V$. 
		In the semiparametric model characterized by the sole restriction 
		\eqref{bisemi}, the efficient score for $\alpha$ under the law $(\talpha,\dot{\eta})$ is:
		\begin{equation*}
		S^*(\talpha; \dot{\eta}) = \omega_{\text{\it eff}}(V) (A-e(V)) (H(\talpha) - p_0(V)), 
		\end{equation*}
		where $\omega_{\text{\it eff}}(V)$ is the efficient weighting function given in the on-line supplementary materials; 
		$H(\alpha) = Y\{\exp(-A\alpha^T W)\}$ when $\theta(V) = \log (\hbox{\rm RR}(V))$, 
		and $H(\alpha) = Y - A \tanh(\alpha^T W)$ when $\theta(V) = \emph{arctanh}(\hbox{\rm RD}(V))$.  
		Further, if for some $\sigma > 0$,
		\begin{equation}
		\label{eqn:positivity}
		e(V), p_0(V) \in (\sigma, 1-\sigma)
		\end{equation}
		with probability $1$, then the efficient score is  bounded.
	\end{theorem}

	The proof is left to the on-line supplementary materials. It follows that the solution $\widehat{\alpha}_{\hbox{\scriptsize\it eff}}$ to
	\begin{equation*}
	\mathbb{P}_n\left[ \omega_{\text{\it eff}}(V) (A-e(V)) (H(\alpha) - p_0(V)) \right]= 0,
	\end{equation*}
	has asymptotic variance equal to the semiparametric variance bound in the model defined by the sole restriction \eqref{bisemi}; here
	$\mathbb{P}_n(\cdot )$ denotes the empirical expectation.

	Of course $\widehat{\alpha}_{\hbox{\scriptsize \it eff}}$ is not feasible as it depends on unknown quantities $e(V)$ and $p_0(V)$.
	When $V$ is high dimensional or has several continuous elements,  common practice is to specify a parametric model for these population quantities. For example, one may specify models  \eqref{bisemi} and \eqref{bibase} and estimate $p_0(V)$ from \eqref{eqn:p0V_RR} and \eqref{eqn:p0V_RD} for the RR and RD, respectively. One may also specify a parametric model for the propensity score $e(V)$, such as
	\begin{equation}
	\label{eqn:model_ps}
	e(V;\gamma) \equiv  P\left(A=1|V;\gamma\right) =\hbox{expit} \left(\gamma^{T}X\right),
	\end{equation}
	where $X=x\left(V\right) $ is a known
	vector function of $V$ and $\gamma$ is a finite dimensional parameter.
	
	Let $\widehat{\gamma }$ solve 
	\[ 
	\mathbb{P}_n\left[ X\left\{
	A-e(V;\gamma) \right\} \right]=0, 
	\]
	and
	$\left(\widehat{\alpha}, \widehat{\beta } \right) $ solve 
	\[
	0=\mathbb{P}_n\left[ \dfrac{\partial}{\partial (\alpha,\beta)} \ell (\alpha, \beta) \right] \equiv \mathbb{P}_n \left[S({\alpha},{\beta})\right].\]
	Note that here $\widehat{\alpha}$ is the maximum likelihood estimator of Section \ref{sec:bivariate_link}. 
	Under suitable regularity conditions \citep{white1982maximum} the estimators
	$\widehat{\alpha}$, $\widehat{\beta}$, $\widehat{\gamma}$ converge in probability to fixed constants $\alpha^*,\beta^*, \gamma^*$ regardless of whether the models  \eqref{bibase} or \eqref{eqn:model_ps} are correct or not. 
	However, $\alpha^*$ may not be equal to the true value of  $\alpha$ if 
	the nuisance model \eqref{bibase} is misspecified.

	Since $\widehat{\alpha}$ may be inconsistent, we instead estimate $\alpha$ by the solution  $\widehat{\alpha}_{\text{\it DR,eff}}$ to the following estimating equation:
	\begin{equation*}
	\mathbb{P}_n\left[U(\alpha; \widehat{\alpha},\widehat{\beta},\widehat{\gamma})\right]=0,
	\end{equation*}
	where
	\begin{equation}
	\label{eqn:ee_unknown-def}
	U(\alpha; \widehat{\alpha},\widehat{\beta},\widehat{\gamma})  =\omega_{\text{\it eff}}(V;\widehat{\alpha},\widehat{\beta},\widehat{\gamma}) 
	\left(A-e(V;\widehat{\gamma}) \right)
	\left(H(\alpha ) -p_0\left(V; \widehat{\alpha},\widehat{\beta} \right) \right).
	\end{equation}
	Under correct specification of model \eqref{bisemi}, $\widehat{\alpha}_{\text{\it DR,eff}}$ is consistent if either the nuisance model \eqref{bibase} or the propensity score model \eqref{eqn:model_ps} is correct. Furthermore, if we replace the efficient weighting function in \eqref{eqn:ee_unknown-def} with any 
	other function $\omega(V)$, under regularity conditions the solution $\widehat{\alpha}_{DR}(\omega)$ to
	\begin{equation}
	\label{eqn:ee_unknown_general}
	\mathbb{P}_n\left[ U(\alpha; \widehat{\alpha},\widehat{\beta},\widehat{\gamma}; \omega) \right] = 0,
	\end{equation}
	where
	\begin{equation}
	\label{eqn:ee_unknown_general_def}
	U(\alpha; \widehat{\alpha},\widehat{\beta},\widehat{\gamma}; \omega) =\omega(V) 
	\left(A-e(V;\widehat{\gamma}) \right)
	\left(H(\alpha ) -p_0\!\left(V; \widehat{\alpha},\widehat{\beta} \right) \right)
	\end{equation}
	also yields a doubly robust estimator. These results are formally described in the following theorem.  The proof is left to the on-line supplementary materials.
	We define the \emph{union model} to be the set of distributions characterized by {\rm (\ref{bisemi})}
	and \emph{either} model {\rm (\ref{bibase})} \emph{or} the propensity model {\rm (\ref{eqn:model_ps})} being correct.
	
	\begin{theorem}
		\label{thm:double_robustness} Under standard regularity conditions and the positivity assumption \eqref{eqn:positivity}, 
		$\talpha$ is the unique solution to the probability limit of \eqref{eqn:ee_unknown_general} in the union model.
		If $\widehat{\alpha}_{DR}(\omega)$ solves \eqref{eqn:ee_unknown_general} then
		$\widehat{\alpha }_{DR}(\omega)$ is a regular and asymptotically linear (RAL) estimator of $%
		\alpha$ under this model. 
		
		Furthermore,  the influence function of $\widehat{\alpha }_{DR}(\omega)$ is
		given by 
		\begin{equation}
		\label{eqn:utilde}   \tau ^{-1}\widetilde{U}(\talpha; {\alpha}^*,{\beta}^*,{\gamma}^*; \omega), 
		\end{equation}
		where
		\[
		\tau = -E\left[ \partial U(\alpha; \alpha^*,\beta^*,\gamma^*; \omega)/\partial \alpha |_{\alpha =\talpha}\right] 
		\] and 
		\begin{flalign*}
		\MoveEqLeft{\widetilde{U}(\talpha; {\alpha}^*,{\beta}^*,{\gamma}^*; \omega)}     \\[12pt] 
		&= U(\talpha; {\alpha}^*,\beta^*,{\gamma}^*; \omega)  - \\[4pt]
		& \quad	E\left[\left. \dfrac{ \partial U(\talpha; {\alpha},{\beta},{\gamma}^*; \omega)}{\partial ({\alpha}^T,{\beta}^T)}\, \right|_{
			\begin{array}{c}\scriptsize
			{\alpha} = \alpha^*\\[-4pt]
			\scriptsize
			{\beta} =  \beta^*
			\end{array}
		} \right] 	 \times 
		\left\{E \left[\left. \dfrac{\partial S({\alpha},{\beta})}{\partial ({\alpha}^T,{\beta}^T)}\,  
		\right|_{\begin{array}{c}\scriptsize
			{\alpha} = \alpha^*\\[-4pt]
			\scriptsize
			{\beta} =  \beta^*
			\end{array}}  \right]
		\right\}^{-1} S(\alpha^*,\beta^*) - \\[12pt]
		& \quad E\left[ \dfrac{\partial U(\talpha; \alpha^*,
			\beta^*, {\gamma}; \omega)} {\partial {\gamma}} \bigg|_{{\gamma} ={\gamma}^*}%
		\right]  \times  \left\{ E\left[ -X\dfrac{ \partial e(V;{\gamma}) }{\partial
			{\gamma}} \bigg|_{{\gamma} =\gamma^*}\right] \right\}
		^{-1}X\left\{ A-e(V;\gamma^*)\right\}.
		\end{flalign*}
	\end{theorem}
	
	The estimator $\hat{\alpha}_{DR}(\omega)$ is only one of many estimators using the weight function $\omega$ that are doubly-robust (i.e.~regular in the union model);  $\hat{\alpha}_{DR}(\omega)$ is arguably the simplest, both conceptually and computationally. In the last decade there has been  an explosion of research that has, in a number of semi-parametric models, developed a methodology for constructing  doubly
	robust estimators that have much better large sample efficiency (at laws outside of the intersection sub-model)  and improved finite sample performance compared to the simplest DR estimators; \citet{rotnitkzy:vansteelandt:2014} give a review. 
	Though these methods could also be applied in our context, we leave this to future work as it is tangential to our goal of demonstrating the utility of our novel nuisance modeling.

	%
	

	\begin{remark}
		\emph{
			We note that double robustness is a useful property only if it is possible for either nuisance model to be correct} a priori. \emph{In other words,  for the double robustness property to be relevant, equation \eqref{eqn:ee_unknown_general} must be used in combination with an (implicit) model for baseline risk that does not suffer from problem (I). Our approach based on the log OP avoids problem (I) due to the variation independence between parameters in  models \eqref{bisemi} and \eqref{bibase}.} 
	\end{remark}

	\begin{remark}
		\label{remark:intersection}
		\emph{
			One can show that the efficient score for the union model is also $S^*(\talpha; \dot{\eta})$ at the submodel where the propensity score  \eqref{eqn:model_ps} is correctly specified. 	Consequently,  the semiparametric variance bound is the same for the union model and the model defined by the sole restriction \eqref{bisemi}  when the propensity score model is correct. In particular, they are the same at the intersection model where both \eqref{bibase} and \eqref{eqn:model_ps}  are correctly specified. See 
			\cite{robins1994estimation} for related results in a general context.}
	\end{remark}

	Regardless of whether or not the union model is correct, the influence function of $\widehat{\alpha }_{DR}(\omega)$ for fixed $\omega$ is given by
	\eqref{eqn:utilde}, replacing $\talpha$ by the probability limit of $\widehat{\alpha }_{DR}(\omega)$, assuming that this exists. 
	Hence, even under mis-specification of both nuisance models, a consistent estimator of the asymptotic variance of $n^{1/2}\left(
	\widehat{\alpha }_{DR}(\omega ) -\talpha \right) $
	is still $\widehat{\tau }^{-1}\widehat{\Sigma }\widehat{\tau }^{-1}$ with
	
	\[
	\widehat{\Sigma }=\mathbb{P}_n\left[ \widehat{\widetilde{U}}(\widehat{\alpha }_{DR}\left(\omega \right); \widehat{\alpha },%
	\widehat{\beta },\widehat{\gamma }; \omega)\widehat{\widetilde{U}}(%
	\widehat{\alpha }_{DR}\left(\omega \right); \widehat{\alpha },\widehat{\beta },\widehat{\gamma }; \omega)^{T}\right], 
	\] 
	where
	$\widehat{\widetilde{U}}$
	is $\widetilde{U}$
	with all expectations replaced by sample averages and $( \talpha,\alpha^*, \beta^*, \gamma^*)$ replaced by $(\widehat{\alpha}_{DR}(\omega),\widehat{\alpha},\widehat{\beta},\widehat{\gamma}) $, and 
	$$\widehat{\tau }=-\mathbb{P}_n\left[ \partial U(\alpha; \widehat{\alpha }, \widehat{\beta },\widehat{\gamma }; \omega)/\partial
	\alpha^T \big|_{\alpha =\widehat{\alpha }_{DR}(\omega) }%
	\right]. $$
	
	\begin{remark}
		\label{remark:ic}
		\emph{ The influence function of $\widehat{\alpha }_{\text{\it DR,eff}}$ can be obtained in two steps: 1) replace $\omega$ in each term of \eqref{eqn:utilde}, $\tau$ and $\widetilde{U}(\talpha; {\alpha}^*,{\beta}^*,{\gamma}^*; \omega)$ with $\omega_{\text{\it eff}}({\alpha},{\beta},{\gamma})$; 2) evaluate \eqref{eqn:utilde} at the point where $(\alpha,\beta,\gamma) = (\alpha^*, \beta^*, \gamma^*)$.	  A consistent estimator of the asymptotic variance of $n^{1/2}\left(
			\widehat{\alpha }_{\text{\it DR,eff}} -\talpha \right) $	 can then be obtained from the influence function analogously to that for $\widehat{\alpha }_{DR}(\omega)$.  Note that  as $\omega_{\text{\it eff}}$ depends on  parameters $\alpha,\beta,\gamma$, the derivatives following equation \eqref{eqn:utilde} should in general include the derivatives of $\omega_{\text{\it eff}}$ with respect to these parameters. However when the union model is true, these derivatives are not required as they do not contribute to the asymptotic variance.}
	\end{remark}
	
	The following theorem states that
	$\widehat{\alpha}_{\text{\it DR,eff}}$ is locally semiparametric efficient in the union model at the intersection submodel. The proof is left to  the on-line supplementary materials. 
	
	\begin{theorem}
		\label{thm:locally_efficient}
		If both the nuisance model \eqref{bibase} and the propensity score model \eqref{eqn:model_ps} are correct, then the variance of $\widehat{\alpha}_{\hbox{\scriptsize \it DR,eff}}$ attains the semiparametric variance bound under the union model.
	\end{theorem}
	
	\begin{remark}
		\label{remark:efficiency-prop-score}
		\emph{ The asymptotic variance of $\widehat{\alpha}_{\hbox{\scriptsize \it DR,eff}}$ based on a {\it correct} log OP model  and  an {\it incorrect}  propensity score model may be smaller than the asymptotic variance (under the same law) with both models correctly specified. On the other hand, the asymptotic variance of $\widehat{\alpha}_{\hbox{\scriptsize \it DR,eff}}$ based on an {\it incorrect} log OP model and a {\it correct}  propensity score model is never smaller than that obtained under correct specification of both models. See 
			\cite{robins2001comments} for related results in a general context.
		}
	\end{remark}
	

	\section{Simulation studies}
	\label{sec:simulations}
	
	In this section, we evaluate the finite sample performance of our proposed procedure.  We generate data from the models \eqref{bisemi}, \eqref{bibase} and \eqref{eqn:model_ps}, where $\talpha = (0, -1)^T, \tbeta = (-0.5,1)^T$ and $\dot{\gamma}=(0.1,-0.5)^T$. The covariates $V$ include an intercept and a uniform random variable generated from $\text{Unif}(-2,2)$; $w(V), z(V)$ and $x(V)$  are identity functions of $V$. 
	
	We also consider scenarios in which the nuisance models,  \eqref{bibase} and \eqref{eqn:model_ps} are misspecified. In particular, instead of using $V$, the analyst uses covariates $V^\dagger$, which include an intercept and an irrelevant covariate also generated from $\text{Unif}(-2,2)$; the misspecified nuisance models  \eqref{bibase} and \eqref{eqn:model_ps} are, respectively, linear and logistic in $V^\dagger$.  
	
	We consider three estimators:
	\begin{itemize}
		\item[{\sf {mle}}:] The maximum likelihood estimator;
		\item[{\sf {drw}}:] The optimally weighted doubly-robust estimator;
		\item[{\sf {dru}}:] The doubly-robust estimator $\widehat{\alpha}_{DR}(\omega)$ with the (naive) weighting function given by $\omega(V)=V$.
	\end{itemize}
	The standard deviation of {\sf mle} is estimated via inverting the Fisher information matrix, and the standard deviations of {\sf drw} and {\sf dru} are estimated using the methods described in, and right before, Remark \ref{remark:ic}.
	
	We  consider four scenarios:
	\begin{itemize}
		\item[\sf{ bth}:] $V$ is used in both the nuisance model  \eqref{bibase} and the propensity score model \eqref{eqn:model_ps};
		\item[\sf{ psc}:]  $V$ is used in the propensity score model \eqref{eqn:model_ps} but $V^\dagger$ is used in the nuisance model  \eqref{bibase};
		\item[{\sf orc}:] $V$ is used in the nuisance model   \eqref{bibase} but $V^\dagger$ is used  in the propensity score model \eqref{eqn:model_ps};
		\item[{\sf bad}:] $V^\dagger$ is used in both the nuisance model    \eqref{bibase} and the propensity score model \eqref{eqn:model_ps}.
	\end{itemize}
	
	Note that because the propensity score is not used in {\sf mle}, 
	results for {\sf mle.orc} are identical to {\sf mle.bth}, and results for {\sf mle.psc} are identical to {\sf mle.bad}. 
	In our simulations we suppose that the model for the target of interest,  \eqref{bisemi}, is correctly specified. 
	%
	%
	All the simulation results are based on $1000$ Monte-Carlo runs of $n=500$ and $n=1000$ units each.
	
	\begin{table}
		\begin{center}
			\caption{Monte Carlo results of the proposed MLE and DR estimators. 
				The true values for $\alpha_0$ and $\alpha_1$ are 0 and -1, respectively, for both RR and RD. The sample size is  500.}
			\bigskip
			\small
			\label{tab:est}
			\begin{tabular}{rccccccc}
				\toprule
				&       \multicolumn{2}{c}	{ Relative Risk} &  \multicolumn{2}{c}	{ Risk Difference} \\
				\cmidrule(r){2-3} \cmidrule(l){4-5}
				& \multicolumn{1}{c}{$\alpha_0$} & \multicolumn{1}{c}{$\alpha_1$} &   \multicolumn{1}{c}{$\alpha_0 $} & \multicolumn{1}{c}{$\alpha_1 $} \\
				\midrule
				\multicolumn{1}{l}{Bias(SE) \quad \quad}	 & & & \\
				{\sf mle.bth} & 0.008(0.004) & -0.021(0.005) & 0.004(0.002) & -0.009(0.003) \\
				{\sf mle.bad} & -0.403(0.004) & 0.021(0.004) & -0.028(0.002) & -0.007(0.003) \\[2pt]
				{\sf drw.bth} & 0.008(0.004) & -0.023(0.005) & 0.004(0.002) & -0.009(0.003) \\
				{\sf drw.psc} & 0.007(0.004) & -0.022(0.005) & 0.003(0.002) & -0.010(0.003) \\
				{\sf drw.lop} & 0.008(0.004) & -0.024(0.005) & 0.004(0.002) & -0.009(0.003) \\
				{\sf drw.bad} & -0.101(0.004) & 0.024(0.005) & -0.029(0.002) & -0.016(0.003) \\
				[2pt]
				{\sf dru.bth} & 0.019(0.005) & -0.057(0.008) & 0.004(0.002) & -0.010(0.003) \\ \\[10pt]
				\multicolumn{1}{l}{SD Accuracy* \quad\quad}	 & & & \\
				{\sf mle.bt}h & 0.991 & 0.968 & 1.015 & 1.017 \\ 
				{\sf mle.bad} & 0.984 & 0.949 & 1.034 & 1.015 \\ [2pt]
				{\sf drw.bth} & 0.996 & 0.971 & 1.015 & 1.007 \\ 
				{\sf drw.psc} & 1.000 & 0.986 & 1.011 & 0.998 \\ 
				{\sf drw.orc} & 0.991 & 0.954 & 1.017 & 1.010 \\ 
				{\sf drw.bad} & 0.946 & 0.953 & 1.023 & 1.010 \\ [2pt]
				{\sf dru.bth} & 0.989 & 0.808 & 1.014 & 0.999 \\ [10pt]
				\multicolumn{1}{l}{Coverage**}	 & & & \\
				{\sf 	mle.bth} & 0.955 & 0.953 & 0.963 & 0.952 \\ 
				{\sf mle.bad} & 0.043 & 0.928 & 0.938 & 0.951 \\ [2pt]
				{\sf 	drw.bth} & 0.958 & 0.956 & 0.960 & 0.940 \\ 
				{\sf 	drw.psc} & 0.961 & 0.963 & 0.960 & 0.944 \\ 
				{\sf 	drw.orc} & 0.960 & 0.946 & 0.960 & 0.953 \\ 
				{\sf 	drw.bad} & 0.845 & 0.935 & 0.938 & 0.952 \\ [2pt]
				{\sf 	dru.bth} & 0.958 & 0.949 & 0.959 & 0.944 \\ 
				\bottomrule 
			\end{tabular}
		\end{center}
		\quad \\[-12pt]
		\footnotesize{*: SD Accuracy = Estimated SD / Monte Carlo SD. \\[-3pt]
			**: Nominal level = 95\%.}
	\end{table}
	
	Table \ref{tab:est} summarizes simulation results.  
	All estimators and scenarios have bias smaller than $0.03$, except for {\sf mle.bad} and {\sf drw.bad}, the latter confirms ``double robustness.'' 
	With both nuisance models correctly specified, the optimally-weighted estimator {\sf drw.bth} has standard deviation smaller 
	than, or comparable  to,  {\sf dru.bth} for both the RR and RD  at both sample sizes, 
	showing that there is indeed an efficiency gain achieved by using the optimal weighting function.  
	In our simulations, results for {\sf dru} under misspecification were very similar to {\sf drw} and hence are omitted.
	We also note that although  $\alpha_0$ and $\alpha_1$ take the same true values in the relative risk  and risk difference model, they have different interpretations. Hence it is not (practically) relevant to directly compare the simulation results across different target parameters (i.e. RR and RD).

	Theory predicts that under correct specification of both the regression model and the propensity score model ({\sf bth}),  
	{\sf mle} achieves the efficiency bound for the parametric outcome regression model, while {\sf drw} achieves the efficiency bound in 
	the larger union model. 
	This is consistent with our simulation results: the standard deviation of {\sf mle.bth} is no larger than that of {\sf drw.bth}.

	We then evaluate the accuracy of the proposed standard deviation estimator by the ratio of the estimated standard deviation and the Monte Carlo standard deviation.  Simulation results show that at modest sample sizes, the estimated standard deviation generally   provides a good approximation to the Monte Carlo standard deviation,  especially when at least one of models \eqref{bibase} and \eqref{eqn:model_ps} is correctly specified. Although the standard deviation estimator associated with {\sf dru.bth} is slightly biased downwards for estimating $\alpha_1$ in the relative risk model  at sample size 500, nonetheless the bias is not sufficient to distort the coverage properties of our interval estimators  and it decreases with sample size. For example, the accuracy of the proposed standard deviation estimator is 0.965 if the sample size is 1000, and 0.980 if the sample size is 10,000. These results are consistent with our claims right after Remark \ref{remark:intersection}. 
	We also 
	compute nominal 95\% Wald-type confidence intervals based on point and standard deviation estimates.  Nominal coverage rates are achieved for all scenarios except for {\sf mle.bad} and {\sf drw.bad}.

	To investigate the sensitivity of the proposed estimator to misspecification of the \emph{functional form} of the nuisance model, we  further consider a simulation scenario where the data are generated from models \eqref{bisemi}, \eqref{eqn:model_ps} and a baseline model $p_0(V) = \gamma^T z(V)$, where $\dot{\alpha} = (0,0.3)^T, \dot{\beta} = (0.5,0.2)^T$ and $\dot{\gamma} = (0.1,-0.5)$.  The covariates $V$  include an intercept and a uniform random variable generated from $\text{Unif}(-1,1)$; $w(V), z(V)$ and $x(V)$  are identity functions of $V$. These coefficient values and covariate ranges are carefully chosen to ensure that the underlying true values of $p_1(V)$ 
	and $p_0(V)$ fall into the unit interval for all possible values of $V$. 
	For comparison, we also fit the true model according to the data generating mechanism; that is, we replace the log odds product model in the proposed approach with a linear baseline risk model.  

	\begin{table}
		\begin{center}
			\caption{Monte Carlo bias and standard error (in parenthesis) with different nuisance models. The data is generated following a linear nuisance model on $p_0$, whereas the proposed approach uses a  linear nuisance model on $\log\text{(OP)}$. 
				The true values for $\alpha_0$ and $\alpha_1$ are 0 and 0.3, respectively for both RR and RD. The sample size is  500.}
			\bigskip
			\small
			\label{tab:est2}
			\begin{tabular}{rccccccc}
				\toprule
				&  Nuisance &       \multicolumn{2}{c}	{ Relative Risk} &  \multicolumn{2}{c}	{ Risk Difference} \\
				\cmidrule(r){3-4} \cmidrule(l){5-6}&
				& \multicolumn{1}{c}{$\alpha_0$} & \multicolumn{1}{c}{$\alpha_1$} &   \multicolumn{1}{c}{$\alpha_0 $} & \multicolumn{1}{c}{$\alpha_1 $} \\
				\midrule
				{\sf mle.bth} & log(OP) & -0.003(0.003) & -0.004(0.005) & -0.002(0.001) & -0.001(0.002) \\ 
				{\sf mle.bth}  & $p_0$ & -0.003(0.003) & -0.003(0.005) & -0.001(0.001) & -0.010(0.002) \\ 
				{\sf mle.bad} & log(OP) & -0.042(0.003) & 0.044(0.005) & -0.062(0.001) & 0.029(0.003) \\ 
				{\sf mle.bad}  & $p_0$ & -0.073(0.003) & 0.357(0.003) & -0.019(0.001) & 0.193(0.001) \\ 
				{\sf drw.bth} & log(OP) & -0.004(0.003) & 0.002(0.005) & -0.002(0.001) & 0.000(0.002) \\ 
				{\sf drw.bth}  & $p_0$ & -0.004(0.003) & 0.003(0.005) & -0.002(0.001) & 0.000(0.002) \\ 
				{\sf drw.psc} & log(OP) & -0.004(0.003) & 0.000(0.005) & -0.002(0.001) & -0.001(0.002) \\ 
				{\sf drw.psc}  & $p_0$ & -0.008(0.003) & 0.000(0.005) & -0.002(0.001) & -0.003(0.002) \\ 
				{\sf drw.orc} & log(OP) & 0.000(0.003) & -0.015(0.005) & -0.002(0.001) & -0.001(0.002) \\ 
				{\sf drw.orc}  & $p_0$ & -0.003(0.003) & 0.002(0.005) & -0.002(0.001) & -0.001(0.002) \\ 
				{\sf drw.bad} & log(OP) & -0.097(0.003) & 0.027(0.005) & -0.061(0.001) & -0.007(0.002) \\ 
				{\sf drw.bad} & $p_0$ & -0.071(0.003) & -0.014(0.005) & -0.040(0.001) & -0.009(0.002) \\ 
				\bottomrule 
			\end{tabular}
		\end{center}
	\end{table}
	
	Table \ref{tab:est2} summarizes the simulation results. 
	Similar to the previous simulation, we consider four  scenarios: {\sf bth}, {\sf psc}, {\sf orc}, {\sf bad}, with the only difference being that the nuisance model can now be either the linear log odds product model or the linear baseline risk model; the latter model is a reasonable alternative here owing to the restricted ranges of the covariates. All findings are compatible with our theoretical development. We note further that in the context of our simulations,
	when $V^\dagger$  is used in both the nuisance model and the propensity score model (i.e. {\sf bad}), our proposed nuisance model is more robust with the maximum likelihood estimator, whereas a linear nuisance model on the baseline risk appears to be more robust with the doubly robust estimator. Also with the proposed $\log$(OP) nuisance model, the estimation results are moderately biased for {\sf drw.orc}. This is because contrary to the data-generating process, the $\log$(OP) model implies a non-linear model   \eqref{eqn:p0V_RR} or \eqref{eqn:p0V_RD}  on the baseline risk. Thus even if we include covariate $V$ in the linear log odds product model, the nuisance model is still misspecified in its functional form, which leads to bias in estimation. Finally, the bias of {\sf drw.orc} with $\log\text{(OP)}$ is smaller compared to the bias of {\sf drw.bad} with $p_0$, suggesting that under our simulation setting, bias from omitting a relevant variable is larger than bias from misspecification of the functional form of the nuisance model.

	\section{Application to fetal data}\label{sec:data-analysis}
	
	We illustrate the proposed novel models  with data from an obstetric study. The data set consists of observations on 14,484 women who delivered at Beth Israel Hospital, Boston from Jan 1970 to Dec 1975. Of these women, 50.4$\%$ received electronic fetal monitoring (EFM). The scientific goal is to evaluate the impact of EFM on 
	cesarean section (CS) rates, which is quantified by the ratio of CS rates for monitored and unmonitored women. It was found previously that to avoid potential confounding bias, it was sufficient to control for 4 variables: nulliparity ({\sf nulli}), arrest of labor progression ({\sf arrest}), breech ({\sf breech}) and year of study ({\sf year}) \citep{neutra1980effect}. As there was extreme heterogeneity across groups defined by these variables, we let the covariates $W$ and $Z$  be identical and include the main terms of these four variables as well as  all two-way and three-way interaction terms of {\sf nulli}, {\sf arrest} and {\sf breech}. The variable {\sf year} was included in the model as a continuous variable. 
	We refer interested readers to  \cite{neutra1980effect} for more details of the study.
	
	
	In our analysis, we applied the following seven methods to the fetal data set: i) {\sf mle}: the proposed RR model fitted using MLE; ii) {\sf dr}: the proposed RR model fitted using DR estimation with the optimal weights; iii) {\sf dr.un}: the proposed RR model fitted using DR estimation, with weight matrix given by the covariates; iv) {\sf poisson}: a Poisson regression model fitted using MLE and robust sandwich standard error estimates; v) {\sf log-binomial}: a log-binomial regression model fitted using MLE;  vi)  {\sf dr.p0}
	DR estimation with a linear nuisance model on the baseline risk, with weight matrix given by the covariates; vii) {\sf logistic}: first fit a logistic regression model; then estimate the conditional RR for a given subgroup from the fitted risks. Models iv), v), vii)  are fitted with the {\tt glm} function in {\tt R}. 

	Figure \ref{fig:fetal} summarizes the  fitted risks from the proposed RR model  in different subgroups.  The parameters were estimated using maximum likelihood estimation.  As expected, the CS rates were very low among women with no complication, but were very high among nulliparas with arrest of labor progression and breech presentation. The effect of EFM was very different  across confounder groups. EFM was associated with a decrease in CS rates in the {\sf breech} only, {\sf nulli}+{\sf breech} and {\sf nulli}+{\sf arrest}+{\sf breech} groups, but was associated with an increase in CS rates in the {\sf arrest} only and {\sf arrest}+{\sf breech} groups. These findings are consistent with the opposing effects reported in \cite{neutra1980effect}.
	
	\begin{figure}
		\begin{center}
			\includegraphics[width=\textwidth]{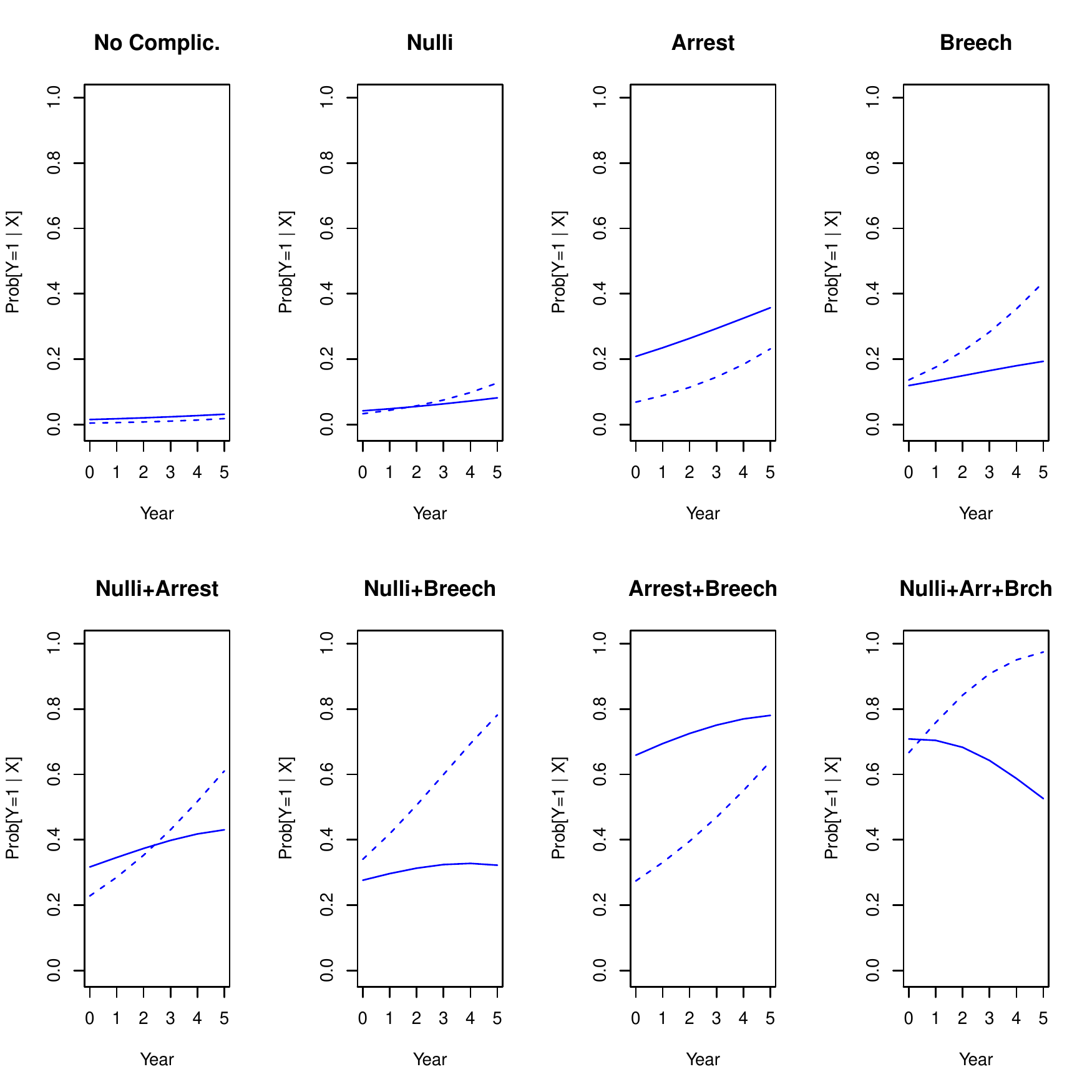}
		\end{center}
		\caption{Estimated risks of cesarean section rates by subgroups.  The dashed lines denote the unmonitored groups, while the solid lines denote the monitored groups. The model parameters are fitted with maximum likelihood.}
		\label{fig:fetal}
	\end{figure}

	Table \ref{tab:real_data} summarizes the comparison between  the proposed RR model and popular GLMs. The bootstrapped estimates are based on 1000 nonparametric bootstrap samples.  The log-binomial regression model failed to give an answer as the program 
	could not
	work out reasonable starting values. The logistic regression model could not be directly used to estimate the dependence of RR on the covariates. 
	The rest of the models gave comparable point estimates. For the proposed estimators, the bootstrapped standard error estimates agree with the proposed standard error estimates except for interaction terms {\sf br*arr} and {\sf arr*br*nul}. Note that only 83 women had both a breech presentation and arrest of labor progression; within them, only 54 women were nulliparas. Hence this discrepancy in standard error estimates is likely due to the small sample size for these subgroups. In contrast, for the Poisson model, the bootstrapped standard errors are substantially smaller than the robust sandwich error estimates for almost all covariates. 
	Furthermore, for interaction terms {\sf br*arr} and {\sf arr*br*nul}, the bootstrapped standard errors of Poisson regression estimates are much larger than those of the proposed RR model. Similarly with the {\sf dr.p0} method.  These suggest that estimates obtained under the proposed RR model may be more stable in small sample settings.


	\begin{table}
		\begin{center}
			\caption{Comparison of model coefficients estimated using different methods}
			\bigskip
			\label{tab:real_data}
			\footnotesize
			\begin{tabular}{rccccccccccccccc@{\;\;\;}r@{\quad\;}}
				\toprule
				& \multicolumn{1}{c}{\scriptsize \sf intercept} & \multicolumn{1}{c}{\scriptsize \sf year} & \multicolumn{1}{c}{\scriptsize \sf nulli} & 
				\multicolumn{1}{c}{\scriptsize \sf arrest} & \multicolumn{1}{c}{\scriptsize \sf breech} &
				\multicolumn{1}{c}{\scriptsize  \sf br*arr} & \multicolumn{1}{c}{\scriptsize \sf arr*nulli} & \multicolumn{1}{c}{\scriptsize \sf br*nulli} 
				& \multicolumn{1}{c}{\scriptsize \sf arr*br*nul} \\ 
				\midrule
				\multicolumn{2}{l}{Point estimate \quad}	  & & &&&&&& \\[2pt]
				{\sf mle} & 1.234 & -0.136 & -0.999 & -0.122 & -1.368 & 1.133 & 0.215 & 0.924 & -0.957\\ 
				{\sf dr} &  1.339 & -0.106 & -1.162 & -0.297 & -1.551 & 1.109 & 0.347 & 1.005 & -0.833 \\ 
				{\sf dr.un} & 1.375 & -0.110 & -1.190 & -0.351 & -1.584 & 1.170 & 0.410 & 1.047 & -0.933 \\ 
				{\sf poisson} & 1.291 & -0.140 & -1.002 & -0.118 & -1.393 & 0.960 & 0.148 & 0.892 & -0.781 \\
				{\sf dr.p0}  &  1.463 & -0.112 & -1.277 & -0.440 & -1.694 & 1.334 & 0.487 & 1.153 & -1.090 \\
				[10pt]
				\multicolumn{2}{l}{Estimated SE*  \quad}	  & &&&&&&& \\[2pt]
				{\sf mle} & 0.229 & 0.029 & 0.239 & 0.344 & 0.340 & 0.597 & 0.371 & 0.396 & 0.656 \\
				{\sf dr} & 0.232 & 0.036 & 0.244 & 0.360 & 0.351 & 0.586 & 0.388 & 0.404 & 0.647 \\
				{\sf dr.un} & 0.243 & 0.037 & 0.251 & 0.356 & 0.351 & 0.583 & 0.385 & 0.406 & 0.646 \\
				{\sf poisson} & 0.912 & 0.167 & 1.069 & 1.024 & 1.008 & 1.288 & 1.359 & 1.332 & 1.660 \\[10pt]
				\multicolumn{2}{l}{Bootstrapped SE  \quad}	  & &&&&&&& \\[2pt]
				{\sf mle} & 0.235 & 0.030 & 0.236 & 0.370 & 0.348 & 0.833 & 0.392 & 0.404 & 0.881 \\
				{\sf dr} & 0.235 & 0.036 & 0.240 & 0.385 & 0.361 & 0.895 & 0.406 & 0.415 & 0.940 \\
				{\sf dr.un} & 0.245 & 0.037 & 0.247 & 0.378 & 0.361 & 0.985 & 0.401 & 0.418 & 1.034 \\
				{\sf poisson} & 0.234 & 0.029 & 0.237 & 0.373 & 0.349 & 1.744 & 0.396 & 0.405 & 1.778 \\
				{\sf dr.p0} & 0.270 & 0.038 & 0.271 & 0.394 & 0.378 & 1.994 & 0.417 & 0.438 & 2.028 \\
				\bottomrule
			\end{tabular}
		\end{center}
		\footnotesize{*: As there is no readily applicable software for computing the standard errors of coefficients estimated with {\sf dr.p0}, we only report the bootstrapped estimates for the standard errors. }
	\end{table}
	
	%

	We further illustrate these methods by examining  a particular subgroup.  As an example, we estimated the relative risk associated with EFM for women who visited the hospital in 1970 and had no complications. In other words, these women are not nulliparas and had normal presentation and labor progression. The  Bootstrap
	method is used for estimating the standard errors and confidence intervals.  One can see from Table \ref{tab:illustration} that all five methods suggest that for this particular subgroup, the CS rate in the monitored group is estimated to be around 3.43 to 4.32 times of that in the unmonitored group. However, the  $95\%$ confidence intervals 
	with the {\sf dr.p0} and {\sf logistic} methods are wider than the others. 

	\begin{table}
		\centering
		\caption{Comparison of inference results on women delivered  in 1970 with no complications}
		\label{tab:illustration}
		\bigskip
		\begin{tabular}{cccc}
			\hline
			& Relative risk estimate & 95\% Confidence interval & P-value \\
			\hline
			{\sf mle} & 3.434 & [2.164,5.447] & $<.001$ \\ 
			{\sf dr} & 3.815 & [2.405,6.051] &  $<.001$  \\ 
			{\sf dr.un} & 3.953 & [2.446,6.391] & $<.001$ \\ 
			{\sf poisson} & 3.635 & [2.298,5.749] &  $<.001$  \\ 
			{\sf dr.p0} & 4.321 & [2.545,7.335] & $<.001$ \\ 
			{\sf logistic} & 4.153 & [1.792,6.515] & 0.009 \\ 
			\hline
		\end{tabular}
	\end{table}

	\section{Discussion}\label{sec:discuss}

	We have presented a general approach to modeling the RR and RD as functions of baseline covariates.
	Our results fill an important methodological gap since 
	many analysts report \OR s as it is simple to 
	model the dependence of the OR on baseline covariates via logistic regression even though many practitioners regard  the RR and RD
	as much more interpretable than \OR s. We have also described methods that are consistent 
	for estimating the RR and RD under correct specification of models for these quantities in conjunction with either a correct model 
	for the propensity score or the odds product.  The code for implementing our methods is available in a new {\tt R} package {\tt brm}  \citep[short for ``binary regression model'',][]{wangbrm}.

	%

	The key methodological insight in this paper is that a non-GLM approach may be used to solve the dilemma between choosing a parameter that is collapsible while wishing to have a model with an unconstrained parameter space. By avoiding the GLM we are able to  separate the target model from the nuisance model.
	
	%
	
	In this paper, we have assumed that the observed data set is a representative sample from the population of interest. 
	The RR and RD can, however, be estimated
	in case-control studies given prior knowledge of the sampling probabilities or estimated from auxiliary data even without a rare disease assumption.  The proposed methods are also related to the literature on modeling the influence of covariates on risk. For example, it is of interest that a referee saw a connection between the approach of \cite{fine1999proportional} who modeled the cumulative incidence function rather than the cause-specific hazard and our approach that modeled the RR or RD rather than the OR.

	\section*{Acknowledgements}
	
This research was supported by U.S. National Institutes of Health grant R01 AI032475, AI113251 and ONR grant N00014-15-1-2672.
	The authors thank Sander Greenland for helpful conversations, encouragement and the data from \cite{neutra1980effect}. The authors also thank John Copas, Dick Kronmal, Eric Tchetgen Tchetgen and Ken Rice for valuable comments.

	\thispagestyle{empty}
	\bibliographystyle{apalike}
	\bibliography{causal}

\clearpage

\begin{center}
	
	{\LARGE Supplementary Materials for ``On Modeling and Estimation for the Relative Risk and Risk Difference''} 
	\vspace{1cm}
	
	{\Large Thomas S. Richardson, James M. Robins  and Linbo Wang} $\ $
	
	
\end{center}
\setcounter{equation}{0}
\setcounter{figure}{0}
\setcounter{table}{0}
\setcounter{page}{1}
\makeatletter
\renewcommand{\theequation}{S\arabic{equation}}
\renewcommand{\thefigure}{S\arabic{figure}}
\setcounter{section}{0}

\section{Plots for implied probabilities from the proposed risk difference models}
\label{appendix:plots}

Figure \ref{fig:fitted_probabilities-rd} plots $p_1(v)$ as functions of $\theta(v)$ and $\phi(v)$, where $\theta(v) = \text{arctanh} \text{RD}(v)$.  Plots with $p_0(v)$ are similar in shapes and hence omitted. Under the proposed models,  the inverse maps (2.4) and (2.5) are sigmoid functions of $\theta(v)$ and $\phi(v)$; these are somewhat similar to those from a logistic regression model. In contrast, as one can see from the lower panels of Figure \ref{fig:fitted_probabilities-rd}, the inverse maps from a linear mean model do not level off even when the probabilities approach 1.

\begin{figure}[!ht]
	\begin{center}
		\begin{minipage}{0.48\textwidth}
			\begin{center}
				\includegraphics[width=\textwidth]{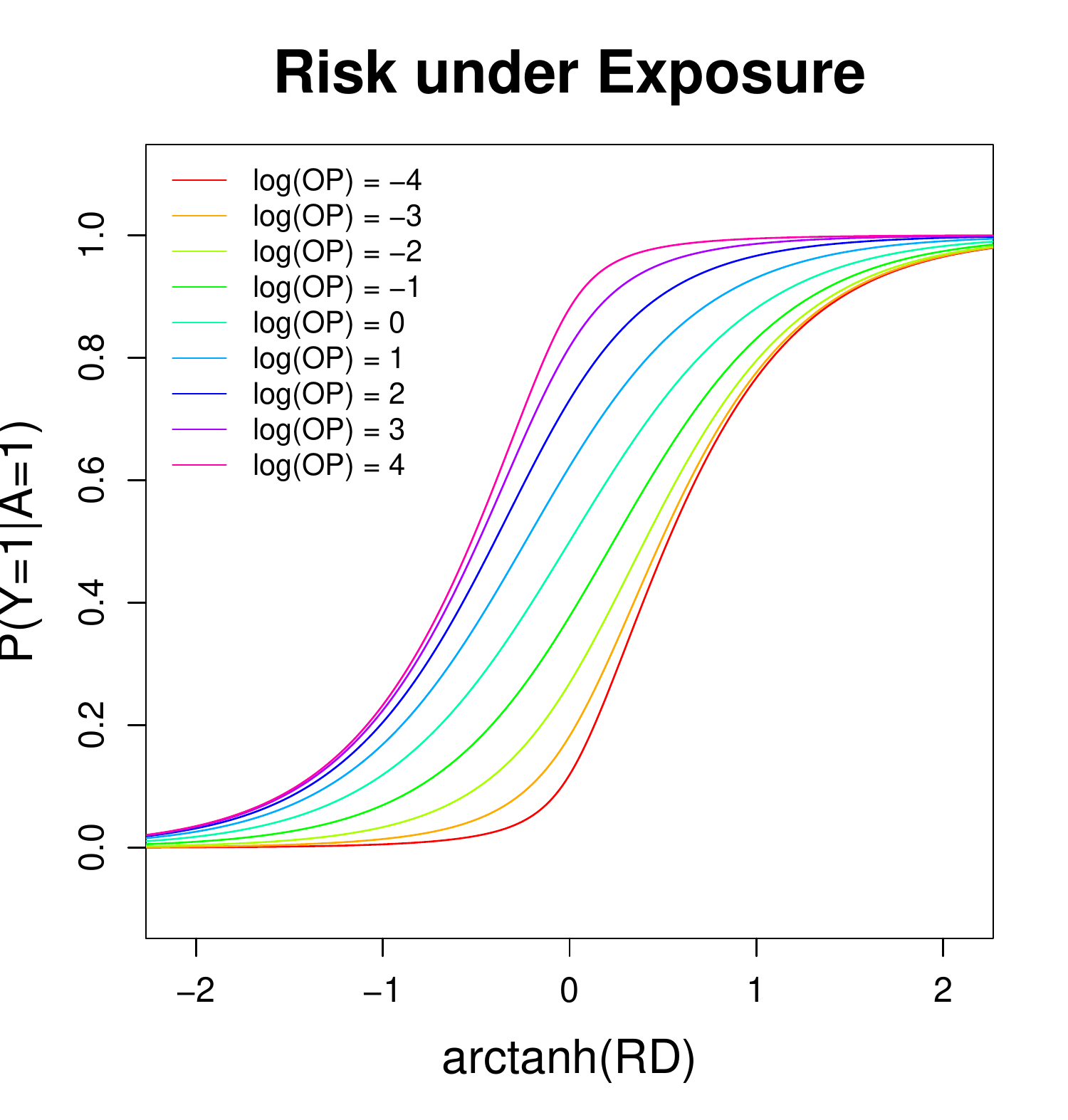}\\
			\end{center}
		\end{minipage}
		\begin{minipage}{0.48\textwidth}
			\begin{center}
				\includegraphics[width=\textwidth]{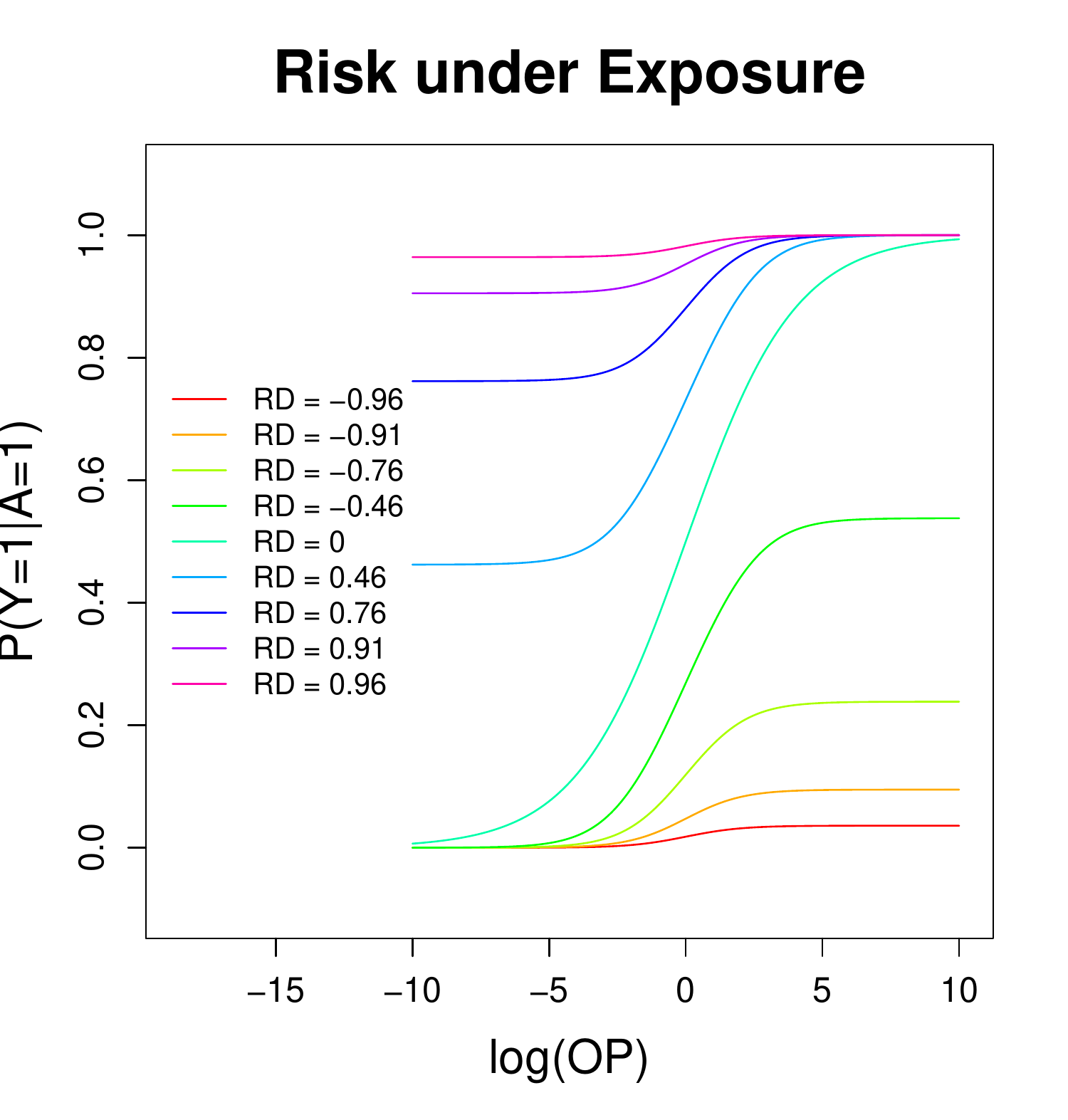}\\
			\end{center}
		\end{minipage}
		\begin{minipage}{0.48\textwidth}
			\begin{center}
				\includegraphics[width=\textwidth]{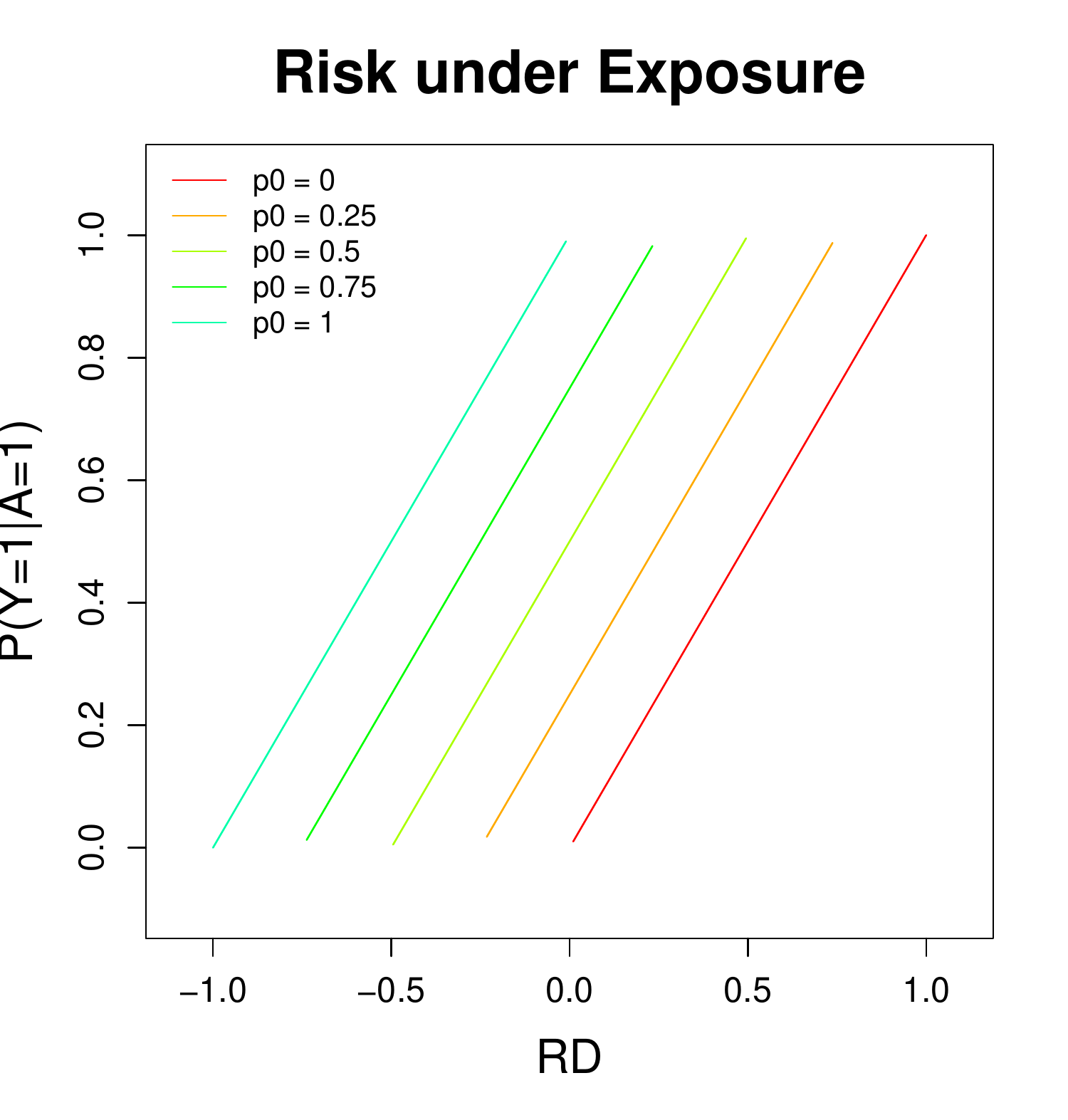}\\
			\end{center}
		\end{minipage}
		\begin{minipage}{0.48\textwidth}
			\begin{center}
				\includegraphics[width=\textwidth]{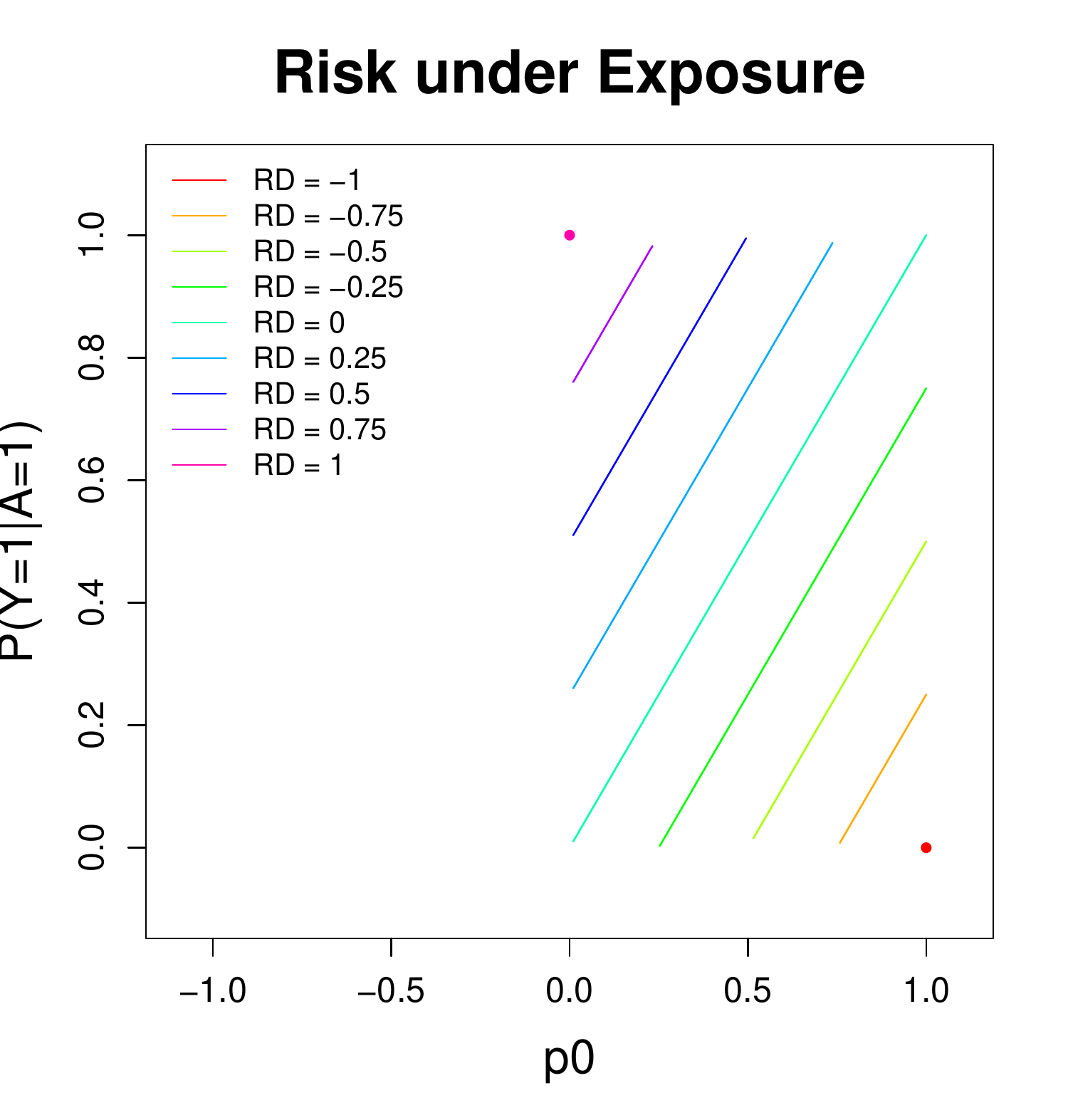}\\
			\end{center}
		\end{minipage}
	\end{center}
	\caption{Implied probabilities from the proposed risk difference model (upper panels) and the linear mean model (lower panels). For the proposed risk difference model, curves for implied risks under non-exposure have the same shapes to those for implied risk under exposure.}
	\label{fig:fitted_probabilities-rd}
\end{figure}

\section{Proof of Theorem 3.1}
\label{appendix_proof_thm1}

\begin{proof}
	\noindent{\it Efficient score for conditional relative risk} \\
	
	\citet[\S A.15]{van2011targeted}  showed that the efficient score for the conditional relative risk is 
	$$
	S^*(\dot{\alpha}, \dot{\eta}) = h^*(A|V) \left( H(\dot{\alpha}) - p_0(V) \right),
	$$
	where 
	\[
	h^*(A|V) = \dfrac{p_A}{p_0(1-p_A)} \left(  AW - 
	\dfrac{E\left[ \dfrac{AW p_A}{1-p_A} \given[\Big] V \right]}{ E\left[ \dfrac{p_A}{1-p_A} \given[\Big] V \right] }   \right),\]
	$p_0 \equiv p_0(V)$ and $p_A \equiv  P(Y=1|A,V)$.
	By analytic calculation,  it is straightforward to verify that 
	$$
	h^*(A|V) = \omega_{\text{\it eff}}(V)(A-e(V)),
	$$
	where 
	\[
	\omega_{\text{\it eff}}(V) = \dfrac{W}{e(V)(1-p_0)} \dfrac{E\left[ \dfrac{Ap_A}{1-p_A} \given[\Big] V \right]}{ E\left[ \dfrac{p_A}{1-p_A} \given[\Big] V \right] } .
	\]
	
	\bigskip
	\noindent{\it Efficient score  for conditional risk difference} \\
	
	The model (2.1) is now given by 
	$$
	p_A  = p_0(V) + \tanh(\theta(V)) A. 
	$$ 
	The score function $S_{\alpha}$ is hence
	$$
	S_{\alpha} = \dfrac{A \rho_{\alpha} (Y-p_A)}{p_A(1-p_A)},
	$$
	where 
	$
	\rho \equiv  \hbox{RD}(V) = \tanh(\theta(V))$ and $\rho_{\alpha} =  \dfrac{\partial}{\partial \alpha} \tanh(\theta(V)) = (1-\rho^2) W$.

	To calculate the tangent space $T_{\eta}$,  we note that it is the direct sum of three orthogonal spaces: $T_{\eta} = T_{p_0} \oplus T_{A|V} \oplus T_{V}$. Moreover, as $\alpha$  depends on the observed data distribution only through $p_{Y|A,V}$, the projection of the score function onto $T_{A|V} \oplus T_{V}$ equals zero. To calculate the tangent space $T_{p_0}$,  we consider submodels centered on $p_0$ and indexed by a scalar $\delta$ (taking values in a neighborhood of zero)
	that are implied by $p_A =  p_0(V) + A \rho+ \delta h(V)$ for an arbitrary function $h(V)$.
	It is easy to see that this indeed implies a submodel obeying (2.1). The score of this submodel at $\delta=0$ equals $h(V)(Y-p_A)/(p_A(1-p_A))$; hence 
	$$
	T_{p_0} = \left\{  \dfrac{  Y-p_A}{p_A(1-p_A)}h(V): h(V) \right\}.
	$$

	To obtain the projection of $S_{\alpha}$ onto $T_{p_0}$, we need to find $h^*(V)$ such that,
	\begin{flalign*}
	E \left[\left\{  u(A,V) (Y-p_A) - \dfrac{h^*(V) (Y-p_A)}{p_A(1-p_A)}  \right\}   \dfrac{h(V) (Y-p_A)}{p_A(1-p_A)}    \right] &= 0, \text{for all\ } h(V),
	\end{flalign*}
	where $u(A,V) = {A\rho_{\alpha}}/({p_A (1-p_A)})$.
	Integrating out $Y$ gives
	\begin{flalign*}
	E \left[  h(V)    \dfrac{ (u(A,V)p_A(1-p_A) - h^*(V) ) }{p_A(1-p_A)}    \right] &= 0, \text{for all\ } h(V).
	\end{flalign*}
	Hence 
	\[
	h^*(V) = \dfrac{E\left[ \dfrac{A\rho_{\alpha}}{p_A(1-p_A)} \given[\Big] V \right]}{ E\left[ \dfrac{1}{p_A(1-p_A)} \given[\Big] V \right] }  = \dfrac{E\left[ \dfrac{A(1-\rho^2)}{p_A(1-p_A)} \given[\Big] V \right]}{ E\left[ \dfrac{1}{p_A(1-p_A)} \given[\Big] V \right] }W.
	\] 
	Consequently, 
	$$
	S^*(\dot{\alpha}, \dot{\eta}) = \dfrac{Y-p_A}{p_A(1-p_A)} \left( A \rho_{\alpha} - 
	\dfrac{E\left[ \dfrac{A\rho_{\alpha}}{p_A(1-p_A)} \given[\Big] V \right]}{ E\left[ \dfrac{1}{p_A(1-p_A)} \given[\Big] V \right] }  \right).
	$$
	By analytic calculation, it is easy to verify that 
	$$
	S^*(\dot{\alpha}, \dot{\eta})  = \omega_{\text{\it eff}}(V) (A-e(V)) (H(\dot{\alpha}) - p_0(V)),
	$$
	where 
	$\omega_{\text{\it eff}} =h^*(V) / (e(V) p_0(1-p_0)) $.
\end{proof}

\section{Proof of Theorem 3.2}
\label{appendix_proof_thm2}
\begin{proof}
	By Slutsky's theorem, it is easy to see that the probability limit of (3.5) is
	\begin{equation}
	\label{eqn:ee_known}
	U(\alpha; {\alpha^*},{\beta^*},{\gamma^*}; \omega)  \equiv \omega(V) 
	\left(A-e(V;{\gamma^*}) \right)
	\left(H\left(\alpha \right) -p_0\left(V; {\alpha^*},{\beta^*} \right) \right)=0.
	\end{equation}
	Let $\tbeta$ and $\dot{\gamma}$ be the true values of $\beta$ and $\gamma$, respectively.
	When model  (2.2) is correct,  $(\alpha^*,\beta^*) = (\talpha,\dot{\beta})$; when model  (3.3) is correct, $\gamma^* = \dot{\gamma}$.
	
	To show that 	$\widehat{\alpha }_{DR}(\omega)$ is a RAL estimator of $%
	\alpha$ in the union model characterized by (2.1)
	and either (2.2) or (3.3), we need to show that under this union model,
	\begin{enumerate}
		\item \eqref{eqn:ee_known} is unbiased for estimating ${\alpha}$;
		\item there exists only one solution to the estimating equation 
		\begin{equation}
		\label{eqn:ee_limit}
		E \left[ U({\alpha}; {\alpha^*},{\beta^*},{\gamma^*}; \omega) \right] = 0.
		\end{equation}
	\end{enumerate}
	The asymptotic linearity of $\widehat{\alpha}_{DR}$ and the corresponding influence function formula then directly follow from \citet[Theorem 5.21]{van2000asymptotic}.

	\bigskip
	\noindent{\it Proof of unbiasedness} \\
	
	Observe that under (2.1),  if $\theta(V) = \log \hbox{\rm RR}(V)$, then
	\begin{flalign*}
	E[H(\dot{\alpha})|V] &= E\left[\left. Ye^{-A \dot{\alpha}^T W} \right| V\right]\\[12pt]
	&= p_0(V) (1- e(V))  +  p_1(V)\left({p_0(V)\over p_1(V)}\right) e(V)\\[12pt]
	&= p_0(V);
	\end{flalign*}
	likewise if $\theta(V) = \text{arctanh}(\hbox{\rm RD}(V))$, then
	\begin{flalign*}
	E[H(\dot{\alpha})|V] &= E\left[ Y-A \tanh (\dot{\alpha}^T W) | V\right]\\[12pt]
	&= p_0(V) (1-e(V)) +  \left(p_1(V) - (p_1(V)-p_0(V))\right)  e(V)\\[12pt]
	&= p_0(V).
	\end{flalign*}
	Similarly, one can show that $E\left[ AH(\dot{\alpha}) \,|\, V\right] = p_0(V)e(V)$.
	
	Now suppose that (2.2) is correctly specified, then $(\alpha^*,\beta^*)=(\dot{\alpha},\dot{\beta})$ and $p_0(V) = p_0(V;\dot{\alpha},\dot{\beta})$. Hence
	\begin{flalign*}
	E\left\{ U(\dot{\alpha}; {\alpha^*},{\beta^*},{\gamma^*}; \omega)  |V  \right\} &= 
	E\left[ \left.\omega(V) (A-e(V;\gamma^*))  (H(\dot{\alpha}) - p_0(V;\dot{\alpha},\dot{\beta}) \,  \right| \, V\right]\\[12pt]
	&= \omega(V) \left\{ E\left[ A H(\dot{\alpha}) - A p_0(V) \,|\, V \right]  - e(V;\gamma^*) E\left[ H(\dot{\alpha}) - p_0(V) \,|\, V  \right]   \right\} \\[12pt]
	&= 0.
	\end{flalign*}

	On the other hand, if (3.3) is correctly specified, then $\gamma^* = \dot{\gamma}$ and $e(V) = e(V;\dot{\gamma})$. Hence
	\begin{flalign*}
	\MoveEqLeft{E\left\{ U(\dot{\alpha}; {\alpha^*},{\beta^*},{\gamma^*}; \omega)  \,|\,V  \right\}}\\[6pt]
	&= 
	E\left[ \omega(V) (A-e(V;\dot{\gamma}))  (H(\dot{\alpha}) - p_0(V;\alpha^*,\beta^*) \,|\, V\right]\\[6pt]
	&= \omega(V) \left\{ E\left[ A H(\dot{\alpha}) - e(V) H(\dot{\alpha}) \,|\, V \right]   - 
	p_0(V;\alpha^*,\beta^*)  E\left[ A - e(V) \,|\, V  \right]   \right\}\;=\; 0.
	\end{flalign*}

	\bigskip
	\noindent{\it Proof of existence and uniqueness of solution to \eqref{eqn:ee_limit}} \\
	
	Let 
	\[
	U(\theta; {\alpha^*},{\beta^*},{\gamma^*}; \omega) \equiv \omega(V) 
	\left(A-e(V;{\gamma}^*) \right)
	\left(H(\theta) -p_0\!\left(V; {\alpha}^*,{\beta}^* \right) \right)
	\]
	where $H(\theta) = Y\{\exp(-A\theta)\}$ when $\theta(V) = \log (\hbox{\rm RR}(V))$, 
	and $H(\theta) = Y - A \tanh(\theta)$ when $\theta(V) = \text{arctanh}(\hbox{\rm RD}(V))$.
	(In other words, $U(\theta; {\alpha^*},{\beta^*},{\gamma^*}; \omega)$ is (3.6)
	with ${\alpha}^TW$ replaced by $\theta$.)
	
	We now show that if either (2.2) or (3.3) is correctly specified, with probability 1, there is at most one value $\theta(v)$ solving
	\begin{equation}
	\label{eqn:ee}
	E\left\{ U(\theta; {\alpha^*},{\beta^*},{\gamma^*}; \omega)  \,|\, V=v  \right\} = 0.
	\end{equation}
	Therefore, assuming  a non-degenerate distribution for  $p_0(V)$, there exists only one $\alpha$ such that $\theta(V) =\alpha^T W$ with probability $1$.
	
	For simplicity, for any given $v$, we write $e^*(v) = e(v;\gamma^*)$ and $p_0^*(v) = p_0(v;\alpha^*,\beta^*)$. 
	Without loss of generality, we assume that with probability $1$, $\omega(v)>0$. If $\theta(v) = \log \hbox{\rm RR}(v)$,  then
	\begin{flalign*}
	\MoveEqLeft{
		\dfrac{\partial}{\partial \theta} E\left[\left. U(\theta; {\alpha^*},{\beta^*},{\gamma^*}; \omega)   \right| V=v\right]}\\
	&= 	\dfrac{\partial}{\partial \theta}  E\left[\left. \omega(V) (A-e^*(V))  (Ye^{-A\theta} - p_0^*(V))\right| V=v \right]\\[12pt]
	&= - \omega(v) E\left[ (A-e^*(v))   AYe^{-A\theta} |V=v \right] \\[12pt]
	&= -C_1(v)  e^{-\theta}  < 0,
	\end{flalign*}
	where $C_1(v) = \omega(v)  e(v) (1-e^*(v))   p_1(v) $ is a positive constant given $v$.
	We  interchange the order of differentiation and integration in the second step as for any value of $\theta$, one can find a bounded open neighborhood 
	$(a,b)$ around it such that  $(A-e^*(v))   AYe^{-A\theta} $ is integrable.  
	This establishes that there is at most one solution to \eqref{eqn:ee}.
	
	To establish existence, we note that 
	\begin{flalign*}
	\MoveEqLeft{\lim_{\theta \rightarrow \infty}E\left[\left. U(\theta; {\alpha^*},{\beta^*},{\gamma^*}; \omega)  \, \right|\, V=v\right]}\\
	&= \lim_{\theta \rightarrow \infty}E\left[\omega(V) (A-e^*(V))  (Ye^{-A\theta} - p_0^*(V) | V=v\right]   \\[12pt]
	&= \omega(v) (1-e(v)) (-e^*(v))  (p_0(v) - p_0^*(v))  + \\
	&\numberthis \label{eqn:whole_1} \qquad \quad \omega(v) e(v) (1-e^*(v))  (- p_0^*(v))     \\[12pt] 
	&= \omega(v) p_0^*(v) (e^*(v)-e(v))  + \\
	&\qquad\quad \omega(v) (1-e(v)) (-e^*(v))  p_0(v).
	\numberthis \label{eqn:whole_2}  
	\end{flalign*}
	If (2.2) is correctly specified, then  the first term in \eqref{eqn:whole_1} vanishes, and the remaining term in \eqref{eqn:whole_1} is negative; likewise if (3.3) is correctly specified, then the first term in \eqref{eqn:whole_2} vanishes, and the remaining term in \eqref{eqn:whole_2} is negative. Hence if  either (2.2) or (3.3) is correctly specified, we have 
	$$
	\lim_{\theta \rightarrow \infty}E\left[\left. U(\theta; {\alpha^*},{\beta^*},{\gamma^*}; \omega)   \right| V=v\right]  < 0.
	$$
	As  $	\dfrac{\partial}{\partial \theta} E\left[\left. U(\theta; {\alpha^*},{\beta^*},{\gamma^*}; \omega)   \right| V=v\right] < -C_1(v)$ for $\theta<0$, we have
	$$
	\lim_{\theta \rightarrow -\infty}E\left[\left. U(\theta; {\alpha^*},{\beta^*},{\gamma^*}; \omega)   \right| V=v\right]  = \infty.
	$$
	By continuity, this establishes the existence of a solution to \eqref{eqn:ee}. 
	
	%
	%
	\bigskip
	
	If $\theta(v) = \text{arctanh}(\text{RD}(v))$,  then
	\begin{flalign*}
	\MoveEqLeft{\dfrac{\partial}{\partial \theta} E\left[\left. U(\theta; {\alpha^*},{\beta^*},{\gamma^*}; \omega)   \right| V=v\right]}\\
	&= 	\dfrac{\partial}{\partial \theta}  E\left[\left. \omega(V) (A-e^*(V))  (Y-A \text{tanh}(\theta) - p_0^*(V)) \right| V=v \right]\\[12pt]
	&= - \omega(v)  E\left[\left. (A-e^*(v))   A(1-\text{tanh}^2(\theta)) \right|V=v \right] \\[12pt]
	&= -C_2(v)  (1-\text{tanh}^2(\theta)) \\[12pt]
	& < 0,
	\end{flalign*}
	where $C_2(v) = \omega(v)  e(v) (1-e^*(v))   $ is a positive constant given $v$. This establishes there is at most one solution to \eqref{eqn:ee}.
	
	To establish existence, we note
	\begin{flalign*}
	\MoveEqLeft{\lim_{\theta \rightarrow \infty}E\left[\left. U(\theta; {\alpha^*},{\beta^*},{\gamma^*}; \omega)   \right| V=v\right]}\\
	&= \lim_{\theta \rightarrow \infty}E\left[\omega(V) (A-e^*(V))  (Y-A\text{tanh}(\theta) - p_0^*(V) | V=v\right]   \\[12pt]
	&= E\left[\omega(V) (A-e^*(V))  (Y-A - p_0^*(V) | V=v\right]   \\[12pt] 
	&= \omega(v) (1-e(v)) (-e^*(v))  (p_0(v) - p_0^*(v))  + \\
	&\numberthis \label{eqn:whole_11} \qquad \quad \omega(v) e(v) (1-e^*(v))  (p_1(v) - 1 - p_0^*(v))     \\[12pt] 
	&= \omega(v) p_0^*(v) (e^*(v)-e(v))  - \\
	&\qquad\quad \omega(v) (1-e(v)) e^*(v)  p_0(v) + \\
	& \qquad \qquad \quad \omega(v) e(v) (1-e^*(v)) (p_1(v)-1).
	\numberthis \label{eqn:whole_12}  
	\end{flalign*}
	If (2.2) is correctly specified, then  the first term in \eqref{eqn:whole_11} vanishes, and the remaining term in \eqref{eqn:whole_11} is negative; likewise if (3.3) is correctly specified, then the first term in \eqref{eqn:whole_12} vanishes, and the remaining term in \eqref{eqn:whole_12} is negative. Hence if  either (2.2) or (3.3) is correctly specified, we have 
	$$
	\lim_{\theta \rightarrow \infty}E\left[\left. U(\theta; {\alpha^*},{\beta^*},{\gamma^*}; \omega)   \right| V=v\right]  < 0.
	$$
	Likewise
	\begin{flalign*}
	\MoveEqLeft{\lim_{\theta \rightarrow -\infty}E\left[\left. U(\theta; {\alpha^*},{\beta^*},{\gamma^*}; \omega)   \right| V=v\right]}\\
	&= \lim_{\theta \rightarrow -\infty}E\left[\left.\omega(V) (A-e^*(V))  (Y-A\text{tanh}(\theta) - p_0^*(V)) \right| V=v\right]   \\[12pt]
	&= E\left[\left.\omega(V) (A-e^*(V))  (Y+A - p_0^*(V) \right| V=v\right]   \\[12pt] 
	&= \omega(v) (1-e(v)) (-e^*(v))  (p_0(v) - p_0^*(v))  + \\
	&\numberthis \label{eqn:whole_21} \qquad \quad \omega(v) e(v) (1-e^*(v))  (p_1(v) + 1 - p_0^*(v))     \\[12pt] 
	&= \omega(v) p_0^*(v) (e^*(v)-e(v))  - \\
	&\qquad\quad \omega(v) (1-e(v)) e^*(v)  p_0(v) + \\
	& \qquad \qquad \quad \omega(v) e(v) (1-e^*(v)) (p_1(v)+1).
	\numberthis \label{eqn:whole_22}  
	\end{flalign*}
	If (2.2) is correctly specified, then  the first term in \eqref{eqn:whole_21} vanishes, and the remaining term in \eqref{eqn:whole_21} is positive; likewise if (3.3) is correctly specified, then the first term in \eqref{eqn:whole_22} vanishes, and the remaining term in \eqref{eqn:whole_22} equals $\omega(v)e(v)(1-e(v)) (1+p_1(v)-p_0(v))$ and is hence positive. Hence if  either (2.2) or (3.3) is correctly specified, we have 
	$$
	\lim_{\theta \rightarrow \infty}E\left[\left. U(\theta; {\alpha^*},{\beta^*},{\gamma^*}; \omega)   \right| V=v\right]  > 0.
	$$
	By continuity, this establishes existence of a solution to \eqref{eqn:ee}. 
\end{proof}

\section{Proof of Theorem 3.6}
\label{appendix:proof_thm4}

\begin{proof}
	If both the nuisance model (2.2) and the propensity score model (3.3) are correct,  then $(\alpha^*,\beta^*) = (\dot{\alpha},\dot{\beta})$ and $\gamma^* = \dot{\gamma}$.
	Hence we have 
	\begin{flalign*}
	\MoveEqLeft{ E\left[ \dfrac{ \partial U(\dot{\alpha};  {\alpha},{\beta},{\gamma}^*)}{\partial ({\alpha}^T,{\beta}^T)} \bigg|_{({\alpha},{\beta}) =  (\alpha^*,\beta^*)} \right] }  \\[12pt]	
	&=  E\left[\omega_{\text{\it eff}}(V)  (A-e(V)) \left(-\dfrac{\partial p_0(V;{\alpha},{\beta})}{\partial ({\alpha},{\beta})} \right) +\right.\\
	&\quad\quad\quad\quad \left. \dfrac{\partial \omega_{\text{\it eff}}(V;{\alpha},{\beta},\dot{\gamma})}{\partial({\alpha},{\beta})}  (A-e(V)) \left(H(\dot{\alpha}) - p_0(V)\right) \bigg|_{({\alpha},{\beta}) =  (\alpha^*,\beta^*)}  \right] =  0
	\end{flalign*}
	and
	\begin{flalign*}
	\MoveEqLeft{ E\left[ \dfrac{U( \dot{\alpha}; \alpha^*,
			\beta^*, {\gamma}) } {\partial {\gamma}} \bigg|_{{\gamma} ={\gamma}^*}%
		\right]} \\[12pt]
	&= E\left[\omega_{\text{\it eff}}(V) \left(-\dfrac{\partial e(V; \gamma) }{\gamma} \right)  \left(H(\dot{\alpha}) - p_0(V)\right)\right. + \\
	&\quad\quad\quad\quad\left.   \dfrac{\partial \omega_{\text{\it eff}}(V;\dot{\alpha},\dot{\beta},\gamma)}{\partial \gamma}  (A-e(V)) \left(H(\dot{\alpha}) - p_0(V)\right)  \bigg|_{{\gamma} ={\gamma}^*} \right]=  0.
	\end{flalign*}
	Following Theorem 3.2, the influence function of $\widehat{\alpha}_{\text{\it DR,eff}}$ is $\tau^{-1} U(\dot{\alpha}; \dot{\alpha},\dot{\beta},\dot{\gamma})$, where 	$\tau = -E\left[ \partial U(\alpha;  \dot{\alpha},\dot{\beta},\dot{\gamma})/\partial \alpha |_{\alpha =\dot{\alpha}}\right] $. From the discussion following Theorem  3.1, it is evident that the variance of $\widehat{\alpha}_{\text{\it DR,eff}}$ attains the bound.
\end{proof}

\end{document}